\documentclass[a4paper,aps,twocolumn]{revtex4}
\usepackage{amssymb}
\usepackage{amsmath}
\usepackage[dvips,xdvi]{graphicx}
\usepackage{epsfig}
\usepackage{amsmath}

\newcommand{\bra}[1]{\langle \, #1 \, |}
\newcommand{\ket}[1]{ | \, #1 \, \rangle }
\newcommand{\bracket}[2]{\langle \, #1 \, | \, #2 \, \rangle}

\newcommand{\id}{\int \! \! d}

\newcommand{\hpsi}{\hat{\psi}}

\newcommand{\beq}{\begin{equation}}
\newcommand{\eeq}{\end{equation}}
\newcommand{\bea}{ \begin{eqnarray} }
\newcommand{\eea}{ \end{eqnarray} }

\begin{document}

\title{Berry's Phases of Ground States
of Interacting Spin-One Bosons: \\
Chains of Monopoles and Monosegments}

\author{Jeroen Wiemer$^{\dagger}$ and Fei 
Zhou$^{\dagger,\dagger\dagger}$}

\affiliation{ITP,Utrecht University, Leuvenlaan 4, 3584 CE Utrecht, The 
Netherlands$^{\dagger}$}
\affiliation{Department of Physics and Astronomy, University of 
British Columbia,\\
6224 Agriculture Road, Vancouver, B.C. V6T 1Z1, Canada$^{\dagger\dagger}$
\footnote{Permanent address}}

\date{\today}

\begin{abstract}
We study Berry's connection potentials of many-body ground
states of spin-one bosons with antiferromagnetic interactions in 
adiabatically varying magnetic fields.
We find that Berry's connection potentials are generally determined by, 
instead of usual 
singular monopoles,
linearly positioned monosegments each of which carries one unit of
topological charge;
in the absence of a magnetic field gradient this distribution of 
monosegments 
becomes a linear chain of monopoles.
Consequently, Berry's phases consist of a series of step
functions of magnetic fields; a magnetic field gradient causes rounding of 
these 
step-functions.
We also calculate Berry's connection fields, profiles of monosegments and show 
that the total topological charge is 
conserved in a parameter space. 
\end{abstract}

\maketitle

\section{Introduction}

In the presence of time-dependent periodical potentials $V(t)=V(t+T)$ 
with a 
period
$T$, an eigenstate of the instantaneous hamiltonian acquires 
a geometric phase known as the Berry's phase\cite{Berry84}.
The Berry's phase is a global characterization of an eigen vector
when it is transported in a Hilbert space.
For a period of $T$ during which
an eigen state $\Psi(V(t))$ evolves adiabatically, the Berry's phase is
simply
$\Phi_B=-{\cal I}m \int_0^T \bra{\Psi} \partial_t\ket{\Psi} 
dt$. 
From
the point of view of fiber bundles, the Berry's phase can be considered as 
the holonomy of a Hermitian fiber bundle; its base space corresponds to a 
parameter space and a fiber is an eigenstate 
which defines a mapping from the parameter space to the Hilbert space. 
This point of view was 
illustrated in \cite{Simon83}.
Berry's phases have been observed in a variety of experiments 
such as NMR and rotation of light polarisation in optical fibres 
etc\cite{Tomita86,Suter87,Tycko87,Shapere89}.

In condensed matter systems where there are large numbers of particles
interacting with each other one way or the other, the subject of Berry's
phases, or more general geometric phases becomes more fascinating and 
intriguing. There are at least two 
interesting aspects of this
subject:
i) The effect of geometric phases on correlations;
ii) The effect of correlations on Berry's phases.

The first issue was addressed in quite a few different occasions. The best
known result perhaps is the geometric phases' effect on spin 
correlations. 
As pointed out a while ago for antiferromagnets, a Berry's phase 
distinguishes integer 
and half-integer
spin chains, or even-integer, odd-integer and half-integer spin 
square lattices and results in different ground states and excitations
\cite{Haldane83,Haldane88,Read89}. The other examples perhaps are the 
possible geometric phases' effects on statistical 
transmutation and fractionalization in spin correlated 
systems \cite{Wilczek83,Wiegmann88,Wen90,Wen91}.
In mesoscopic magnetic systems effects of Berry's phases on 
collective quantum tunneling were investigated
\cite{Bogachek92,Delf92,Loss92,Prokofev93,Tupitspyn97,Barbara99};
quantum interferences of Berry's phases were observed in molecular 
magnets\cite{Wernsdorfer99}.

In this article we are going to address the second aspect, correlations'
effect on Berry's phases. We study Berry's connection potentials 
(one-form), Berry's connection fields (two-form) and 
topological charge densities for ground states of spin-one bosons with 
antiferromagnetic interactions.
In a parameter space 

\begin{equation}
\{ X_a\}, a=1,2,...M,
\end{equation}
the connection potential ${\bf A}_a$, two-form connection fields
${\bf F}_{ab}$, and topological 
charge-current densities $\rho$, ${\bf J}$ are
defined as

\begin{eqnarray}
&& {\bf A}_a(\{ X_c\}; \Psi)=\frac{i}{2}\{ <\Psi|\frac{\partial 
\Psi}{\partial X_a}>-<\frac{\partial \Psi}{\partial X_a}|\Psi> \},
\nonumber \\
&& {\bf F}_{ab}(\{ X_c\};\Psi)=i \{
<\frac{\partial \Psi}{\partial X_a}|\frac{\partial
\Psi}{\partial X_b}>-<\frac{\partial \Psi}{\partial 
X_b}|\frac{\partial \Psi}{\partial X_a}> \}   , 
\nonumber \\
&& 4\pi \rho= \frac{1}{2} \epsilon^{abc}\partial_a {\bf F}_{bc}, \;
4 \pi {\bf J}_a= \partial_b {\bf F}_{ab}.
\end{eqnarray}

Consider Berry's connection potentials, two-form connection fields and 
topological
charges when N spin-one bosons are in a magnetic field 
${\bf B}=(B_x, B_y, B_z)$.
For ground states of N-noninteracting spin-one bosons, one easily 
confirms
that the two-form connection fields are monopole-like due to an N-fold 
magnetic monopole at the point of ${\bf B}=0$ in the parameter space of
$\{B_a\}$, $a=1,2,3$ or $x, y, z$. And the topological current is zero.
That is

\begin{eqnarray}
&& {\bf b}_a=\frac{1}{2} \epsilon_{abc} F_{bc}=Q({\bf B}) 
\frac{{\bf B}_a}{|{\bf B}|^3}, Q({\bf B})=N
\nonumber \\
&& \rho=Q({\bf B})\delta({\bf B}), {\bf J}=0.
\end{eqnarray}
We should emphasize here that the topological charge is localized at 
the origin ${\bf B}=0$ in the parameter space.
Note that the Berry's two-form field ${\bf b}$
is a function of external magnetic fields ${\bf B}$. 

For N spin-one bosons with antiferromagnetic interactions, we are going to
show that the profile of topological charge density is 
a linear chain of
monopoles. In the presence of a field gradient, we also find nonsingular 
monosegments
where topological charges distribute smoothly along certain direction and 
topological 
current flows. And the density profile is a linear chain of monosegments.
Details of 
topological charge density profile depend on the even-odd parity of N, 
magnetic
field gradient and spin relaxation.

The plan of this article is as follows. In section II, we introduce 
the 
system
which interests us and describe the algebras needed for this 
investigation.
In section III, we calculate the Berry's connection fields of $N$ 
interacting spin-one
bosons in a uniform magnetic field.
We show that antiferromagnetic interactions in general expel the 
topological charges outward 
from the origin and result in a linear chain distribution of
monopoles in the
parameter space. 
In section IV, we study the effects of the spin nonconserving process;
we calculate the monosegment profile in the presence of a 
magnetic field gradient. 
In addition, we address 
the Landau-Zener effect.

In both section III and IV,
we show that Berry's phases are suppressed because of 
antiferromagnetic interactions and as functions of 
magnetic field 
consist of a
series of step-like functions. 
Furthermore, the
shape of steps is determined by a magnetic field gradient. We note 
that
in the absence of interactions, the many-body Berry's phase is simply the
sum of each individual spin-one bosons and doesn't depend
on magnetic fields.

In section V we outline an
alternative description of spin-one bosons using quantum rotor models. 
In the effective representation, we
show that the problem of interacting spin-one bosons in a
magnetic field gradient is equivalent to a quantum rotor coupled to a 
quadrupole field.

\section{The microscopic hamiltonian}

We consider spin-one bosons in an optical trap in the dilute 
limit
defined by $n \, a^3 \, \ll 1$ where $a$ is the s-wave scattering length and
$n$ the density.
The hamiltonian is given as
\begin{eqnarray}
\lefteqn{H=\int \! \! d {\bf r} \, \left\{ \, \frac{\hbar^2}{2m} \,
\mbox{\boldmath $\nabla$}
\hat{\psi}_{\alpha}^\dag({\bf r})
\cdot \mbox{\boldmath $\nabla$} \hat{\psi}_{\alpha}({\bf r}) + U({\bf r}) \, 
\hat{\psi}_{\alpha}^\dag({\bf r})
\hat{\psi}_{\alpha}({\bf r}) \right.}  \\
& & \mbox{} + \gamma \, {\bf B} \, \cdot \hat{\psi}^\dag_{\alpha}({\bf r}) \, 
{\bf S}_{\alpha \beta} \,
\hat{\psi}_{\beta}({\bf r}) + \left. \frac{c_0}{2} \, \hat{\psi}_{\alpha}^\dag({\bf r}) 
\hat{\psi}_{\alpha'}^\dag({\bf r})
\hat{\psi}_{\alpha'}({\bf r}) \hat{\psi}_{\alpha}({\bf r}) \right. \nonumber \\ 
&+& \left. \frac{c_2}{2} \,  \hat{\psi}_{\alpha}^\dag({\bf r})
\hat{\psi}_{\alpha'}^\dag({\bf r}) \, {\bf S}_{\alpha \beta} \cdot 
{\bf S}_{\alpha' \beta'} \, 
\hat{\psi}_{\beta'}({\bf r})
\hat{\psi}_{\beta}({\bf r}) \right\}. \nonumber
\end{eqnarray}
Interactions between atoms are approximated by spin-dependent
contact interactions.

In a single mode approximation,
the creation and annihilation operators are defined as
\beq
\hat{\psi}^\dag_{\alpha}({\bf r})= 
\hat{\psi}^{\dag}_ {\alpha} \, \chi_{0}^\ast({\bf r}),
\eeq
$\alpha=m=0,\pm 1$.
$\chi_{0}({\bf r})$ is the lowest orbital mode 
and  $\ket{1,m}$ is a spin-one state with $S_z=m$;
${\bf S}_{\alpha \beta}$ are three matrices;
\begin{eqnarray}
S^x= \frac{1}{\sqrt{2}} \left(
\begin{array}{ccc}
   0 & 1 & 0 \\
      1 & 0 & 1 \\
         0 & 1 & 0
\end{array}\right)&&
S^y= \frac{1}{\sqrt{2}} \left( \begin{array}{ccc}
	     0 & -i & 0 \\
	     i& 0 & -i \\
	     0 & i & 0
     \end{array}\right)\\
S^z=\left( \begin{array}{ccc}
	   1 & 0 & 0 \\
	   0 & 0 & 0 \\
	  0 & 0 & -1
\end{array}\right).&&\nonumber
\end{eqnarray}

The dilute limit of
spin-one bosons with antiferromagnetic interactions
was studied in a few experiments in 
\cite{Stamper98,Stenger98} and also theoretically investigated in various
works \cite{Ho98,Law98,Ohmi98,Ho00}. A 
geometric-description-based nonperturbative approach 
to strong-coupling limits was proposed in \cite{Zhou01,Zhou01a}. In 
optical lattices, correlated
Mott states of spin-one bosons were studied in a series of recent 
papers \cite{Demler02,Zhou02,Imambekov03,Snoek03}; spin nematic, spin 
singlet and dimerized-valence-bond crystals were found for high 
dimensional and low dimensional optical lattices. Unconventional spin 
disordered condensates in homogeneous limits were proposed in various 
papers\cite{Zhou01a,Liu02}; many properties of spin singlet condensates
were further explored in low dimensional optical 
lattices\cite{Zhou03}. In fast rotating traps, correlated quantum liquids
of spin-one bosons were also investigated \cite{Ho02,Schoutens02}.

It is convenient to introduce the following creation-annihilation 
operators
\begin{subequations}
\begin{eqnarray}
\hat{\psi}_{x}^\dagger &=& \frac{1}{\sqrt{2}} ( \hpsi_{ -1}^\dagger -
\hpsi_{1}^\dagger ) \\
\hpsi_{y}^\dagger &=& \frac{i}{\sqrt{2}} ( \hpsi_{-1}^\dagger
+ \hpsi_{1}^\dagger ) \\
\hpsi_{z}^\dagger &=& \hpsi_{0}^\dagger
\end{eqnarray}
\end{subequations}
In this representation total spin operators are defined as
\beq
\hat{{\bf S}}^\alpha=
\hat{\psi}^\dag_\beta {\bf S}^\alpha_{\beta\gamma} \hat{\psi}_\gamma,
{\bf S}^\alpha_{\beta\gamma}=-i \epsilon_{\alpha \beta \gamma} ;
\eeq
$\hat{\psi}^\dag_\alpha$, $\hat{\psi}_{\alpha}$ (from now on, $\alpha=x,y,z$) 
are usual bosonic operators 
obeying the
following commutation relations:
\begin{equation}
\lbrack \hpsi_{\alpha}, \hpsi_{\beta} \rbrack =
\lbrack \hpsi_{\alpha}^\dagger, \hpsi_{\beta}^\dagger \rbrack = 0, \;\;
\lbrack \hpsi_{\alpha}, \hpsi_{\beta}^\dagger \rbrack = 
\delta_{\alpha \beta}.
\end{equation}

Using these results, the hamiltonian can be written in the following form
\bea
H 
&=& (\epsilon_0-g_0-2g_2) \hat{N} + g_0 \, \hat{N}^2 \\ 
&+& g_2 \, \hat{{\bf S}}_{tot}^2 + \gamma {\bf B} \cdot \hat{{\bf S}}_{tot}
\nonumber 
\eea
with
\bea
g_0 &=& \frac{c_0}{2} \int \! \! d {\bf r} \, |\chi_0({\bf r})|^4 \\
g_2 &=& \frac{c_2}{2} \int \! \! d {\bf r} \,  |\chi_0({\bf r})|^4. 
\nonumber
\eea

In terms of singlet pair creation operators
$\hat{A}$,
\beq
\hat{A}^\dag \ket{0} = \frac{1}{\sqrt{6}} 
\left( -2 \hat{\psi}^\dag_{-1} \hat{\psi}^\dag_1 + \hat{\psi}^\dag_0 \hat{\psi}^\dag_0
\right) \ket{0},
\eeq
the total spin operator can also be written as
\beq
\hat{{\bf S}}^2_{tot}=\hat{N}(\hat{N}+1)-6 \, \hat{A}^\dag \hat{A},
\eeq
with 
$\hat{N}=\hat{\psi}_{\alpha}^\dag \hat{\psi}_\alpha$ the number 
operator.
Consequently, we rewrite the Hamiltonian as
\bea
H &=& (\epsilon_0-g_0-g_2) \hat{N} + (g_0+g_2) \hat{N}^2 \\ 
&-& 6 \, g_2
(\hat{A}^\dag \hat{A})  
+ \gamma {\bf B} \cdot \hat{\bf S}_{tot} \nonumber.
\eea

\subsection{Spectrum}

The Hamiltonian commutes with the number operator, squared magnitude of the
total spin operator $\hat{{\bf S}}^2_{tot}$
and operator $\hat{A}^\dag \hat{A}$.
Eigenstates of the hamiltonian are simultaneous eigenstates of the number
operator $\hat{N}$ and operator $\hat{A}^\dag \hat{A}$. 
If $\ket{N,N_s}$ is a state with a {\it total} number of $N$ particles and 
$N_s$ pairs of
singlets, then we have
\beq
\hat{{\bf S}}_{tot}^2 \ket{N,N_s} = (N-2N_s)(N-2N_s+1) \ket{N,N_s}
\eeq
or $S=N-2N_s$.

For N particles in a magnetic field ${\bf B}$ along $z$-axis,
the spin Hamiltonian is 
\beq
H = g_2 \hat{{\bf S}}^{\, 2}_{tot} + \gamma {\bf B} \cdot \hat{{\bf S}}_{tot}.
\eeq
Obviously,
an external magnetic field along the $z$-axis splits the $2S+1$-fold 
degeneracy of states 
with given 
spin $S$ and  ${\bf S}^z$ remains to be a good quantum number.
The energy of an eigen state $|S, S_z=m>$ is

\begin{equation}
E_{S,m}=g_2 \, S(S+1) + \gamma \, B \, m.
\end{equation}
The lowest energy state for a given spin $S$ is $\ket{S,-S}$.
The low energy spectrum at different magnetic fields can be found in FIG.1.

\subsection{Levelcrossings}

\subsubsection{N odd}
Levels $|S_1,m_1>$ and $|S_2, m_2>$ cross at

\beq
B=\frac{g_2}{\gamma} \, \frac{S_2(S_2+1)-S_1(S_1+1)}{m_1-m_2}.
\eeq
For the following values of magnetic fields  

\beq
B_k=\frac{g_2}{\gamma} \, (4 k+5)
\eeq
with $k=0,1,2, \ldots ,M_N  $
where $M_N=\frac{N-3}{2}$ for odd $N$.
we have levelcrossings in
groundstates.

\subsubsection{N even}

For an even N, the levelcrossings in the groundstate take place at
\beq
B_k=\frac{g_2}{\gamma} (4k+3),
\eeq
with $k=0,1,2, \ldots, M_N$ where $M_N=\frac{N-2}{2}$ for even $N$.

\begin{figure}
\setlength{\unitlength}{1mm}
\begin{picture}(0,0)
\put(0,-50){a)}
\put(0,-100){b)}
\end{picture}
\epsfig{file=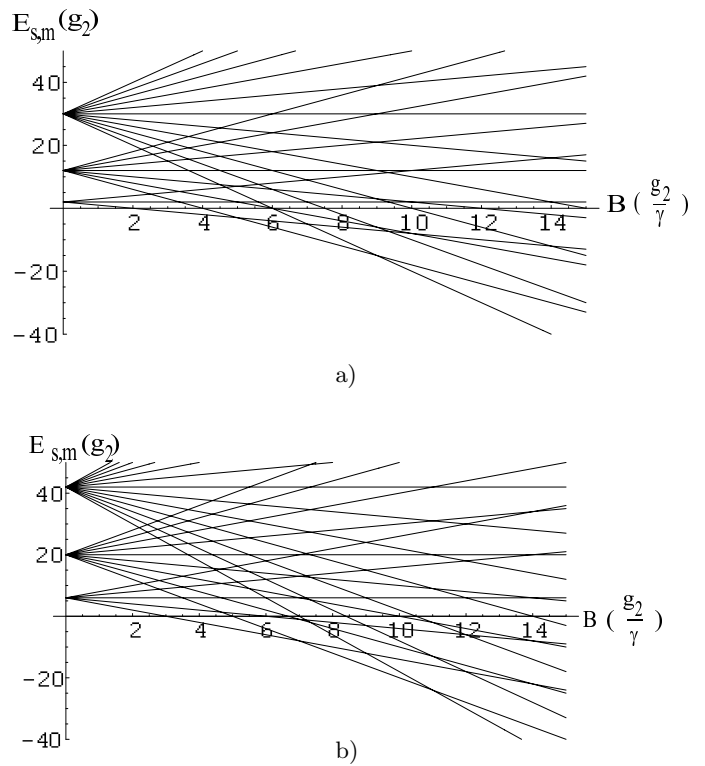,height=100mm,width=90mm}
\caption{Energy spectrum as a function of magnetic field. $E_{s,m}$ is 
given in
units of $g_2$, $B$ in units of $g_2 \, \gamma^{-1}$.
a) is for an odd $N$ and b) is for an even $N$.}
\end{figure}



\section{Monopoles and Berry's connection fields of $N$ interacting 
spin-$1$ 
bosons in homogeneous magnetic fields}
\subsection{Local connection fields and topological charge densities}

The microscopic many-body state of
$\ket{S,S_z = -S}$ 
is given as
\bea
\ket{S,S_z=-S} &=& 
C \, (\hat{\psi}_{-1}^\dag)^S  \hat{A}^{\frac{N-S}{2}} \ket{0}. \nonumber
\eea
Here 
$\hat{A}$ is the singlet creation operator defined before;
$C$ is a normalization factor 
\beq
C= 6^{\frac{N-S}{4}} \, \frac{1}{\sqrt{f(\frac{N-S}{2},S)}}
\eeq
with
\beq
f(M,S)=S! \, M! \, 2^M \, \frac{(2M+2S+1)!!}{(2S+1)!!}.
\eeq

In a spherical coordinate system,
we denote a magnetic field ${\bf B}$, of magnitude $B$ pointing 
in the direction of 
${\bf n}=(\cos \phi \, \sin \theta, \sin \phi \, \sin \theta, \cos \theta)$ as
\beq
{\bf B}=(B, \theta,\phi).
\eeq
For a magnetic field ${\bf B}$ pointing at ${\bf n}$ direction,
ground states 

\beq
|g>= \Psi_{S(B),-S(B)}({\bf n}) 
\eeq
are states
$\ket{S,S_z=-S; {\bf n}}$ with the $z$-axis defined along ${\bf n}$-direction.
The corresponding 
many-body wave functions are

\begin{widetext}
\bea
\Psi_{S(B),-S(B)}({\bf n}) = C  
\left[ -i \sin^2(\frac{\theta}{2}) \, e^{-i \phi} \, \hat{\psi}_{+1}^\dag
+ \frac{i}{\sqrt{2}} \sin \theta 
\,  \hat{\psi}_0^\dag - i \cos^2(\frac{\theta}{2}) \, 
\, e^{i\phi} 
\hat{\psi}_{-1}^{\dag}
\right]^S  
\hat{A}^\dag{}^\frac{N-S}{2} \ket{0}.  
\eea
\end{widetext}

And finally
$S$ is a function of $B$, the magnitude of external magnetic fields.
As shown in FIG.1, 

\beq
S=S_0(N)+2\sum_{k=0,1,2...}^{M_N}\Theta(|{\bf B}|-B_k)
\eeq
$S_0(N)$ is zero for an even $N$ but is unity for an odd $N$.

For these correlated ground states,
Berry's connection potentials in spherical coordinates $(B, \theta, \phi)$ 
can be
defined as 

\bea
&& {\bf A}({\bf B})=-{\cal I}m 
<\Psi_{S,-S}( {\bf n})|\frac{\partial}{\partial B} |\Psi_{S,-S}({\bf 
n})>{\bf e}_B
\nonumber \\
&& -\frac{1}{B}{\cal I}m 
<\Psi_{S,-S}( {\bf n})|\frac{\partial}{\partial {\theta}} 
|\Psi_{S,-S}({\bf 
n})>{\bf e}_\theta \nonumber \\
&&- \frac{1}{B\sin\theta} {\cal I}m 
<\Psi_{S,-S}( {\bf n})|\frac{\partial}{\partial {\phi}} 
|\Psi_{S,-S}({\bf 
n})>{\bf e}_\phi.
\eea

A direct calculation of ${\bf A}$ yields desired results

\bea
&& {\bf A}=-
\frac{\cos\theta}{\sin\theta} \frac{Q(|{\bf B}|)}{|{\bf B}|}{\bf e}_\phi;
\nonumber\\
&& Q(|{\bf B}|)=q_0(N)+ 2\sum_{k=0,1,2...}^{M_N}\Theta(|{\bf B}|-B_k)
\eea
$q_0(N)=1$ for an odd number of particles and $q_0(N)=0$ for an even 
number of
particles.
The two-form connection fields are

\beq
{\bf b}={Q(|{\bf B}|)}\frac{1}
{|{\bf B}|^2}{\bf e}_B +\frac{1}{B}\frac{\partial Q(|{\bf B}|)}{\partial 
B}\frac{\cos\theta}{\sin\theta} {\bf e}_\theta.
\label{b-field}
\eeq

It is worth emphasizing that the radial component of ${\bf 
b}$-fields can be attributed to multiple shells, each of which carries 
exactly two units of charges and is located at $B=B_k$. Indeed, if one 
defines 
$4\pi \rho_B=\nabla \cdot ({\bf 
b}\cdot {\bf e}_B){\bf e}_B$,
one obtains

\begin{equation}
\rho_B= q_0 (N)\delta({\bf B})+\frac{1}{2\pi}
\sum_{k=0,1,2...}^{M_N}\delta(B-B_k)\frac{1}{B^2}
\end{equation}
which indicates an isotropic topological charge distribution of a series 
of shells at $B=B_k$.

The topological charge due to the $\theta$-component
of ${\bf b}$-fields on the other hand consists of two contributions:
a) an isotropic charge distribution of spherical shells exactly
identical to $-\rho_B$ given above; b) linearly distributed monopoles
located at ${\bf B}=\pm B_k{\bf e}_z$ each of which also carries one unit
charge. As a result,
the total amount of topological charge 
due to $\theta$-component 
vanishes identically
on each shell; 
however, the $\theta$-component deforms the isotropic 
distribution on each shell completely.

Finally, the total topological charge and current densities 
are
\begin{widetext}
\bea
&& \rho
=q_0\delta({\bf B})+ \frac{1}{2\pi\sin\theta|{\bf 
B}|^2}\sum_{k=0,1,...}^{M_N} 
\delta(|{\bf B}|-B_k)[\delta(\theta-\pi)+\delta(\theta)]\nonumber \\
&& {\bf J}=\frac{1}{4 \pi |{\bf B}|}  
\frac{\partial^2 Q(|{\bf B}|)}{\partial B^2} 
\cot(\theta) {\bf e}_{\phi}
\label{td}
\eea
\end{widetext}


\begin{figure}
\setlength{\unitlength}{1mm}
\begin{picture}(0,0)
\put(38,2){a)}
\put(38,-70){b)}
\put(38,-140){c)}
\end{picture}
\epsfig{file=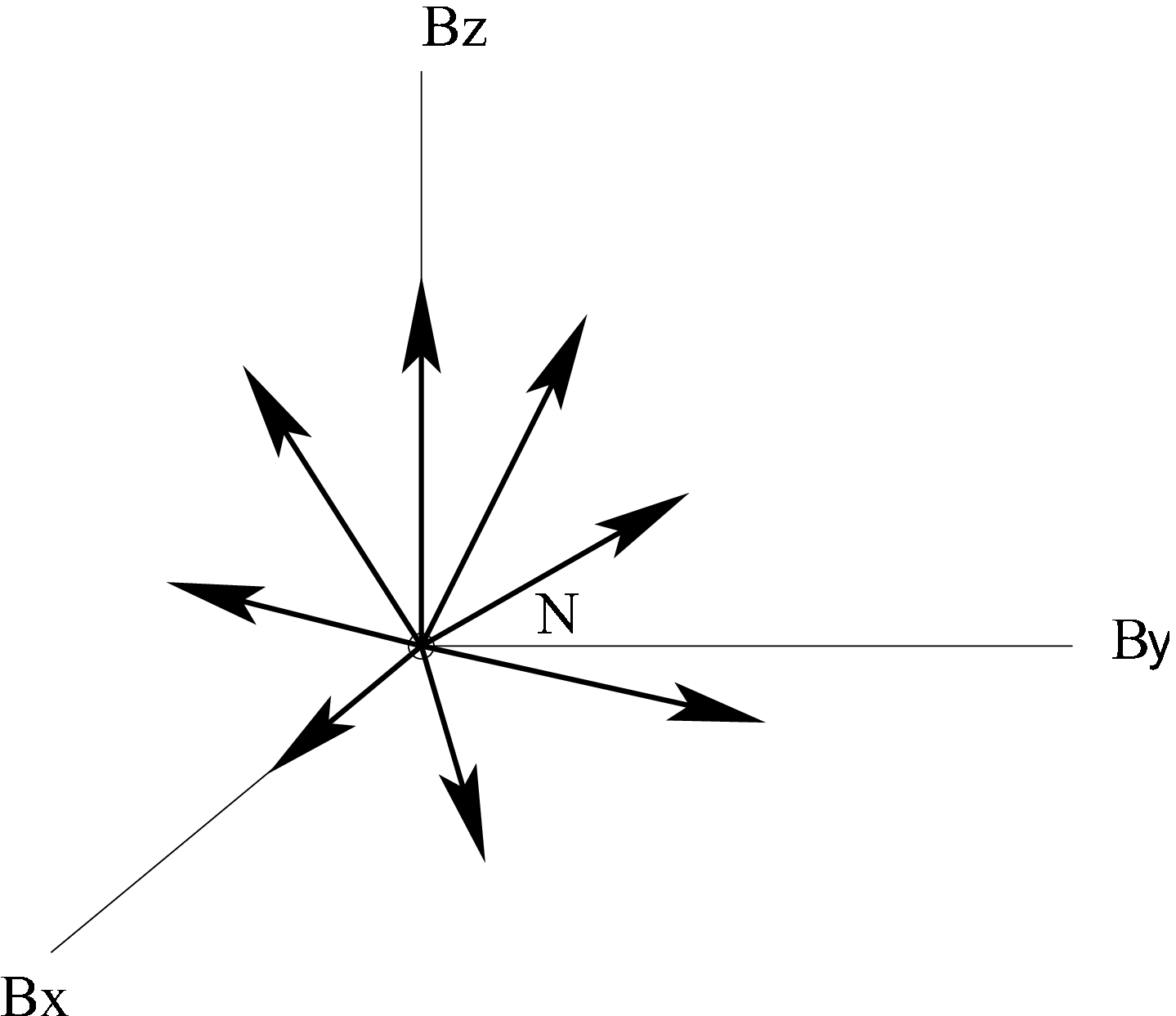,height=70mm,width=70mm}
\epsfig{file=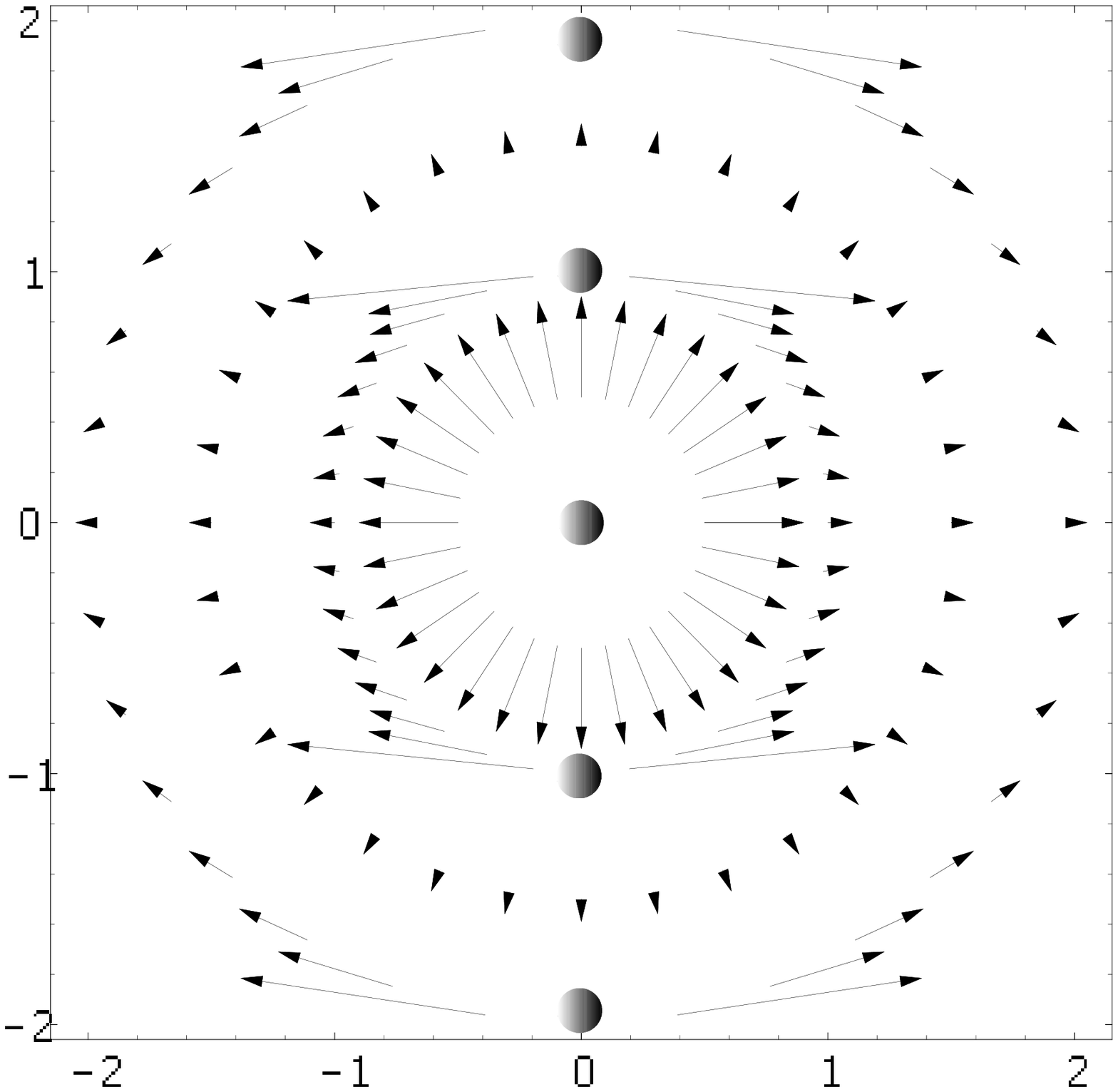,height=70mm,width=70mm}
\epsfig{file=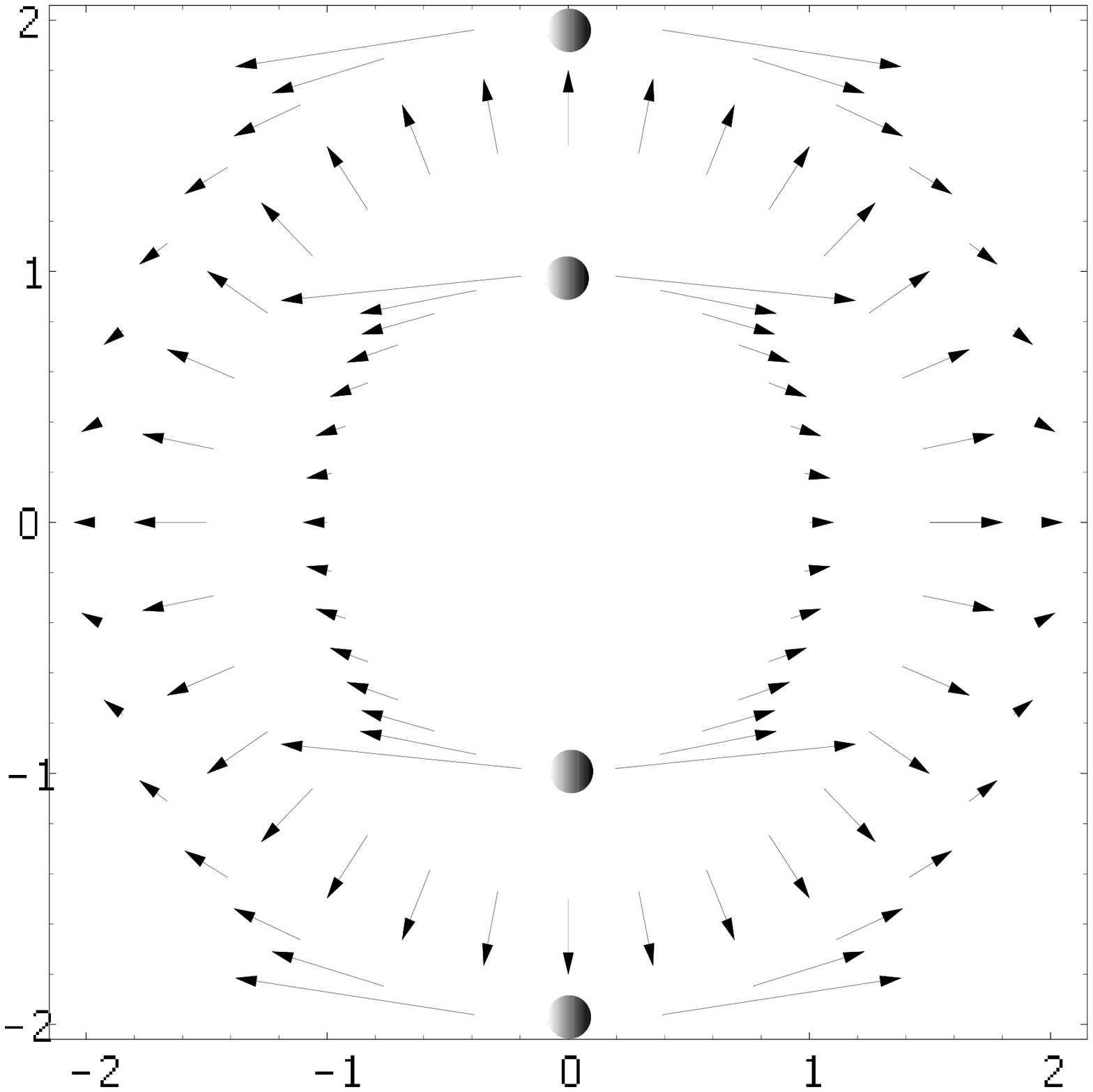,height=70mm,width=70mm}
\caption{Schematic of topological field and charge distribution in the
parameter space $(B_x, B_y, B_z)$.
In a) we show the topological field and charge distribution for 
$N$-noninteracting spin-one bosons.
In b) and c) we show the field and charge distribution for an odd and
even number of spin-one bosons with antiferromagnetic interactions 
respectively;
chains of monopoles are shown here explicitly. 
The ${\bf b}$-fields only have radial component 
except on shells of $B=B_k$, $k=0,1,2...M_N$. 
Note that
monopoles in b) and c) are also distributed on a series of shells with
radius $B=B_k$; and for the even case,
there are no monopoles at the origin
of the parameter space. }
\end{figure}

Compared with noninteracting systems, we find following three new features
in Berry's connection fields and topological charges.

{\bf i)}
Following Eqs.\ref{b-field},\ref{td}, we find that the total 
topological charge within the 
$k$th shell bounded by an outer surface $B_1=B_k+h$ and an inner surface 
$B_2=B_k-h$ where
$B_{k+1}-B_k \gg h$ is a 
conserved integer 
independent of
$B$ and index $k$. Indeed,

\beq
\frac{1}{4\pi}\oint_{S_1}{\bf B} \cdot {\bf e}_B 
dS-\frac{1}{4\pi}\oint_{S_2}{\bf B} \cdot {\bf e}_B  dS
=2. 
\eeq
Here the integration is over two surfaces $S_{1,2}$ of $B=B_{1,2}$.

Therefore each shell defined above still carries over all two units of 
topological charges.
If we introduce the total charge of a surface with radius $B$ as

\beq
\sigma(B)=2\pi \int \rho(B, \theta) B^2 \, \sin \theta \, d\theta,
\eeq
we find

\beq
\sigma(B)=q_0(N) \delta(B) +2\sum_{k=0,1,...}^{M_N}\delta(|{\bf B}|-B_k).
\label{sc}
\eeq
At last, topological currents on each shell circulate around 
the $z$-axis.

As shown in the introduction, for noninteracting spin-one bosons,  
the topological charge is located at the center of the parameter space.
Following Eqs.\ref{td},\ref{sc}, it is evident that antiferromagnetic 
interactions
between spin-one bosons lead to expulsion of topological charges from 
the origin ${\bf B}=0$. As a result, topological charges distribute on 
different shells
with radius $B_k$, $k=0,1,...M_N$; each shell carries two units of 
charges.

However,
following Eq.\ref{td}, one also confirms that
the total topological charge $Q_T$ is conserved; 
$Q_T$ is exactly the number of particles in the many-body state, 
independent of interaction strength. That is

\begin{equation}
Q_T=\int \rho({\bf B}) d{\bf B}=N.
\label{conser}
\end{equation}
One can easily show that the solutions in Eqs.\ref{sc},\ref{conser} are 
independent of
the specific class of ground state wave functions chosen for this 
investigation, though 
the results in Eq.\ref{td} do depend on choices of wave functions.

{\bf ii)}
Furthermore, on each shell the distribution of ${\bf b}$-fields 
breaks the rotational symmetry, unlike in non-interacting cases; 
consequently the charge distribution on each shell is highly anisotropic.
The expulsion of topological charges for the particular set of states 
studied here is along the $z$-axis only.
One can verify that in the case of homogeneous magnetic fields
two units of charges on the $k$th shell are located at $\theta=0,\pi$,
$B=B_k$ points; that is, all charges on each shell are carried by  
two monopoles located at
the northern and southern poles of shells.
So the density profile is a chain of monopoles located at
${\bf B}=\pm B_k{\bf e}_z$, $k+0,1,..M_N$.

This structure however is not generic. As we will show in 
the next section in the presence of a field gradient, charges on
each shell are carried by smooth structures of monosegments instead of 
monopoles.

{\bf iii)}
On each 
shell of $B=B_k$, $k=0,1,2...M_N$ as shown in FIG.2,
the ${\bf b}$-field has a new component along $\theta$-direction. 
This is due to the level crossing or more precisely presence of
different spin states in exact ground states.
The new component represents a Berry's phase 
when a closed path in the parameter space crosses $B=B_k$ surfaces. We will 
come back to this point in the next subsection.

\subsection{Berry's phases I: rotating fields}

When a magnetic field ${\bf B}$ with a given magnitude rotates around 
the z-axis, a correlated state of spin-one atoms 
evolves along a path ${\cal C}$ shown in FIG.3a). During each period
the state acquires a many-body
Berry's phase. 
Consider a rotating magnetic 
${\bf B}(t)=B \, {\bf n}(t)$ with
${\bf n}(t)=( \cos \phi(t) \sin \theta, \sin \phi(t) \sin \theta, \cos 
\theta)$;
$\theta$ is a constant and $\phi=2\pi \Omega t$ is time-dependent.

\begin{figure}
\epsfig{file=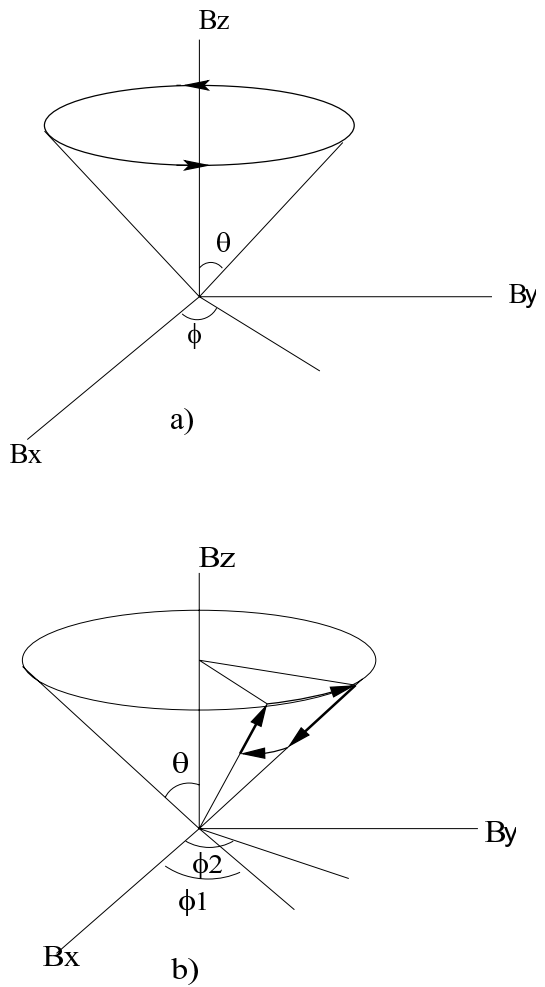,height=130mm,width=70mm}
\caption{Paths in a parameter space. a) A path corresponding to
a rotating field; b) A path for a modulating field discussed in this
subsection; the path is in $\theta=\theta_0$ surface.}
\end{figure}

Previous results on connection fields indicate
that Berry's phases of many-body states 
\beq
\Phi_B({\cal C})=\oint_C {\bf A}\cdot d{\bf B}
\eeq
depend on the magnitude of magnetic 
fields. 
Following calculations in Appendix C,
\beq
\Phi_B =-2\pi  \, \cos \theta \, Q(B).
\label{Bp}
\eeq
As shown in Eq.\ref{Bp}, at a given magnetic field which is much smaller 
than $B_{M_N}$, the Berry's 
phase is always strongly 
suppressed
because of antiferromagnetic correlations.
We summarize our results in FIG.4.

The non-analytical behavior of Berry's phases in rotating fields
is consistent with an anisotropic distribution of monopoles discussed in
the last section. 
Consider two infinitesimal paths ${\cal C}_{1,2}$ centered 
at and also oriented along the $z$-axis.
${\cal C}_1$ is slightly above $-B_k{\bf e}_z$ and ${\cal C}_2$
slightly below. The Berry's phases evaluated in this way are singular at 
points $-B_k {\bf e}_z$ and experience a jump

\beq
\Phi_B({\cal C}_2)-\Phi_B({\cal C}_1)=4\pi.
\eeq
Thus the step-function plotted in FIG.4 indeed implies that the surface 
bounded by path 
$C_1$ and $C_2$ enclose a monopole at ${\bf B}=-B_k{\bf e}_z$.
One can apply similar argument at ${\bf B}=B_k {\bf e}_z$ and arrive
at the same conclusions.

\begin{figure}
\label{fig:Bbp}
\epsfig{file=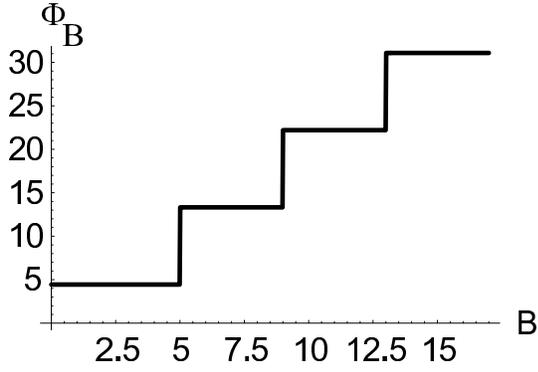}
\caption{A Berry's phase for
a rotating field with $\theta=\frac{3 \pi}{4}$ and various $B$; $B$ is 
given
in $g_2 \gamma^{-1}$. $N$ is taken to be an odd number.}
\end{figure}

\subsection{Berry's phases II: modulating fields}

For non-interacting cases, ${\bf b}$-fields only have ${\bf 
e}_B$-components.
If a path completely lies in a plane parallel to ${\bf e}_B$,
a quantum state doesn't acquire a Berry's phase because 
the topological flux threading the path should be zero for an obvious 
reason.

One of interesting aspects of Eq.\ref{b-field}
is that there is a $\theta$-component in ${\bf b}$-fields because of
antiferromagnetic interactions. This 
yields
a new possibility to study Berry's phases which is uniquely associated
with interacting particles.
Consider a path that lies in a plane of $\theta=\theta_0$ and is bounded 
by

\beq
B=B_{1,2}, \phi=\phi_{1,2}.
\eeq
as shown in FIG. 3b).

Following Eq.\ref{b-field}, one obtains

\begin{equation}
\phi_B=-\cos\theta_0 \, (\phi_2-\phi_1) [Q(B_2)-Q(B_1)].
\end{equation}
This Berry's phase is nonzero only when $B=B_{1,2}$ surfaces are at two 
sides of $B=B_k$ surfaces.
For $B_{k+1} > B_2$ $> B_k >$ $ B_1 > B_{k-1}$ and $\theta_0=3\pi/4$,
$\phi_B=(\phi_1-\phi_2)\sqrt{2}$.
The Berry's phase for modulating fields of this kind vanishes identically 
for noninteracting particles.

Discussions on geometrical phases are valid when 
slow spin relaxation is allowed so that the system can always reach true 
ground states in different
magnetic fields within practically relevant time intervals.
Furthermore we also assume that the quantum symmetry restoring
time is much shorter than measurement time; the issue of symmetry 
restoring of condensates of spin-one bosons was addressed in some detail 
in Ref.\cite{Zhou03}.

\section{Monosegments and Berry's phases of $N$ interacting spin-$1$
bosons in inhomogeneous magnetic fields}

\subsection{Spin conservation and magnetic field
gradient}

In a homogeneous magnetic field,
because of total spin conservation there is no mixing 
between states with different spins.
However, a 
gradient in
magnetic fields, as we will see doesn't conserve total spin and
does mix states with different spins. This
results in "level repulsion" when two spin states approach each 
other.

To show that an inhomogeneous magnetic field violates the conservation of
total spin, we consider the commutator $[H,\hat{S}_{tot}^2]$ with $H$ 
given by
\bea
&& H= g_2 \, \hat{{\bf S}}_{tot}^{\, 2} + \gamma \int \! \! d {\bf x} \,
{\bf B}({\bf x}) \cdot
\hat{\psi}^\dag_\alpha {\bf S}_{\alpha \beta} \hat{\psi}_\beta,
\nonumber \\
&& {\bf S}^\alpha_{\beta\gamma}=-i\epsilon^\alpha_{\beta\gamma}.
\eea

One easily verifies that
\bea
\lefteqn{\left[ \int \! \! d {\bf x} \, \, {\bf B}({\bf x}) \cdot 
\hat{\psi}_{\alpha}^\dag
{\bf S}_{\alpha \beta} \hat{\psi}_{\beta} , \hat{\bf S}^2_{tot} 
\right]}   \\
&&= -(2i) (\int \! \! d {\bf x} \, {\bf B}({\bf x}) \times
\hat{\psi}^\dag_\beta
{\bf S}_{\beta \beta'} \hat{\psi}_\beta') \nonumber \\ 
&& \cdot (\int \! \! d {\bf y} \, 
\hat{\psi}^\dag_\alpha({\bf y})
{\bf S}_{\alpha \alpha'} \hat{\psi}_{\alpha'}({\bf y})) \nonumber   \\
&& -2 \int \! \! d {\bf x} \,\,  \hat{\psi}_\alpha^{\dag} ({\bf x}) \, 
{\bf B}({\bf x})
\cdot {\bf S}_{\alpha \beta} \, \hat{\psi}_\beta({\bf x}).  \nonumber
\label{ct}
\eea
This commutator only vanishes in a homogeneous magnetic field but
is nonzero when a field gradient is present(see Appendix B).

\subsection{Mixing of many-body states with different total spins}

For spatially varying magnetic fields ${\bf B}({\bf x})$,
we define new creation and annihilation operators $\hat{\psi}'_\alpha$,
$\hat{\psi}'_\alpha{}^{\dagger}$ in a local triad where the local magnetic
field points at the $z$ direction; 
the local field operators can be obtained by the following spin rotation

\beq
\hat{\psi}'_\alpha({\bf x})= U_{\alpha \beta}({\bf x}) \, 
\hat{\psi}_\beta({\bf x}),
\eeq
with $U_{\alpha \beta}({\bf x})$ given by
\beq
U_{\alpha \beta}({\bf x})=\exp (i \, \mbox{\boldmath $\theta$}({\bf x}) 
\cdot {\bf S}_{\alpha \beta}).
\eeq

The Zeeman splitting in terms of spin-rotated operators is:
\beq
H_{Zeeman}=\gamma \int \! \! d {\bf x} \, B({\bf x}) \, \hat{\psi}'_\alpha{} ^{\dag}
S^z_{\alpha \beta} \hat{\psi}'_\beta \, , 
S^z_{\alpha\beta}=-i\epsilon^z_{\alpha\beta}.
\eeq
with $B({\bf x})=|{\bf B}({\bf x})|$.
On rotated spin-one fields, external magnetic fields always act along
$Z$-axis.

One also finds that
the spin-dependent and spin-independent interaction 
terms are invariant under 
the local spin rotation. 
However, the kinetic energy transforms nontrivially 
and acquires a new term, $H_{kin} \rightarrow H_{kin }+H_1$ (see 
Appendix C).

Consider a field distribution
\beq
{\bf B}({\bf x})=B_0(1-G'z) {\bf \hat{z}} +G' B_0 \, x  \, {\bf \hat{x}}.
\label{fd}
\eeq
We assume that $G'$ is very small compared to the dimension of the system,
i.e. $G' \, \Omega^{\frac{1}{3}} \ll 1$. This ansatz was employed for
the study of field gradient effects\cite{Ho00}.
$\mbox{\boldmath $\theta$}$, which is determined by the relative 
orientation of magnetic
fields with respect to the z-axis, can be calculated as 
\beq
\mbox{\boldmath $\theta$}({\bf x})={\bf \hat{z}} \times {\bf \hat{B}}({\bf x})  
=G' \, x {\bf \hat{y}}.
\eeq

Following Appendix C, we find that
the spin rotation considered above effectively orients external fields 
along
the $z$-axis and results in a 
new term 
$H_1$ in the single mode 
hamiltonian, i.e.

\beq
H [{\bf B}({\bf x})]\rightarrow H[B({\bf x}){\bf e}_z]+H_1
\eeq
and

\beq
H_1 =
\frac{\epsilon}{2} \sum_{\alpha\neq y}
\hat{\psi}^{\dag}_{\alpha}
\hat{\psi}_{\alpha}  ,
\label{H1}
\eeq
where $\epsilon=\frac{\hbar^2 G'^2 }{m}$.

The tunneling matrix elements 
can be calculated as


\bea
\bra{1,-1} \hat{H}_1 \ket{3,-3} &=& \frac{\epsilon}{2} \sqrt{\frac{3}{70}}
\sqrt{(N-1)(N+4)} \nonumber \\
&\stackrel{N \rightarrow \infty}{=}& 
\frac{\epsilon}{2} \sqrt{\frac{3}{70}} N  
\eea

For general level crossings in ground states, we obtain the following
matrix elements. For an odd $N$,
\bea
\lefteqn{\Delta_k =\bra{(2k+1),-(2k+1)} \hat{H}_1 \ket{(2k+3),-(2k+3)}} \\  
&&= \frac{\hbar^2 G'^2}{4m}
\sqrt{\frac{(2k+2)(N-2k-1)(2k+3)(N+2k+4)}{(4k+5)(4k+7)}}. \nonumber
\label{meo}
\eea
And for an even $N$,

\bea
\lefteqn{\Delta_k = \bra{2k,-2k} \hat{H}_1 \ket{2k+2,-2k-2}} \\
&&=\frac{\hbar^2 G'{}^2}{4m}
\sqrt{\frac{(2k+1)(N-2k)(2k+2)(N+2k+3)}{(4k+3)(4k+5)}.} \nonumber
\label{mee}
\eea

Therefore a field gradient in magnetic fields
has nonvanishing matrix elements between states with different spins and 
leads to mixing of corresponding 
many-body states. 
We will study the resultant spectrum in the next subsection.

\subsection{An effective Hamiltonian close to crossing points $B_k$}

For values of
$|{\bf B}|$ close to the level crossing points $B_k$, we have a small 
tunneling 
term
calculated above. If two nearly degenerate states are far away from
other levels, the 
Hilbert space can be truncated into a two-level
Hilbert space and the effective Hamiltonian in this subspace can be
easily determined.

When $B$ is close to $B_k$, we therefore obtain an 
effective Hamiltonian in the truncated space

\beq
H=\left( \begin{array}{cc}
     E_{S_k,-S_k}(B) & \Delta_k \\
     \Delta_k & E_{S_k+2,-S_k-2}(B)
     \end{array} \right).
\eeq
$S_k=2k+1$ for an odd $N$ and $S_k=2k$ for an even $N$.And once again,
$k=0,1,2...M_N$.
We have assumed that the gradient is small and $\Delta_k \ll g_2$ so that 
the truncation discussed here is always applicable
when $B$ is close to $B_k$.

For two levels $|1,-1>$ and $|3,-3>$
at $B\sim B_0=\frac{5 g_2}{\gamma}$,

\beq
H=\left( \begin{array}{cc}
     E_{1,-1}(B) & \Delta_0 \\
     \Delta_0 & E_{3,-3}(B)
     \end{array} \right)
\eeq
with
\beq
\Delta_0 = \sqrt{\frac{3}{70}} \, \frac{\hbar^2 G'^2 }{2m} \, N. 
\eeq

$E_{1,-1}=E_{3,-3}=E_0$ with
$E_0=-3 g_2$ when $B=B_0$.
Introduce
$\delta B= B-B_0$, $\delta E_{l,-l}=E_{l,-l}-E_0$. One obtains
$\delta 
E_{1,-1}=-\gamma \delta B$, $\delta E_{3,-3}=-3 \gamma \delta B$.
The eigen values can be expressed as
\beq
E_{\pm}(\delta B)= 
-3 g_2 -2 \gamma \, \delta B
\pm ( \gamma^2 \delta B^2 + \Delta_0^2)^{\frac{1}{2}}. 
\eeq

\subsection{connection fields of spin mixed many-body states}

Because of the field gradient, matrix elements of the Hamiltonian 
between states of different spins are nonzero.
Eigenstates are generally 
superpositions of states with different spins.

Close to level crossing points $B_k$,  in the truncated Hilbert space the 
eigen states are 

\bea
&& \Psi({\bf n})=\delta_{k1} \,\ket{S_k,-S_k; {\bf n}} + \delta_{k2}\, 
\ket{S_k+2,-S_k-2;{\bf n}}, \nonumber \\
&& |\delta_{k1}|^2 +|\delta_{k2}|^2=1;
\eea
the corresponding many-body
microscopic wave functions are

\begin{widetext}
\beq
{\Psi({\bf n})}= \delta_{k1} \, C \, (\hat{\psi^{''}}^\dag_{-1} )^{S_k}\, 
(\hat{A}^\dag)^{\frac{N-S_k}{2}}
\ket{0} \, + \, \delta_{k2} \, D (\hat{\psi^{''}}^\dag_{-1})^{S_k+2} \,
(\hat{A}^\dag)^{\frac{N-S_k-2}{2}} \ket{0},
\eeq
\end{widetext}
where $C$ and $D$ are normalization constants.
$\hat{\psi^{''}}^+_{-1}$ is a creation operator defined in a local
frame discussed in Appendix C1.  
For a given $k$,

\bea
S_k=2k+1 \, \mbox{for an odd $N$}\nonumber \\
S_k=2k \, \mbox{for an even $N$} 
\eea

Finally, as the magnetic field increases from below $B_k$ to above $B_k$,
$\delta_{k1}$ varies from one to zero and $\delta_{k2}$ from zero to one.
Taking into account the effective Hamiltonian in the truncated Hilbert 
space derived in section IV C, we obtain the following field dependence
of the coefficients $\delta_{k1,k2}$ when $B$ is close to $B_k$
\bea
\delta_{k1} &=& \frac{ (\gamma \delta B_k-\sqrt{\Delta_k^2+\gamma^2 \delta
B_k^2})}{\sqrt{2(\gamma^2 \delta B_k^2+\Delta^2_k-\gamma \delta B_k
\sqrt{\Delta^2_k+\gamma^2 \delta B_k^2})}} \\
\delta_{k2} &=& \frac{\Delta_k}{\sqrt{2(\gamma^2 \delta B^2_k + 
\Delta_k^2-\gamma
\delta B_k \sqrt{\Delta_k^2+\gamma^2 \delta B_k^2})}}. \nonumber
\label{co}
\eea
In Eq.\ref{co}, we have introduced $\delta B_k=B-B_k$.
$\Delta_k$ ($k=0,1,2...M_N$) represent the matrix elements calculated in 
Eq.51;
$\Delta_k=\bra{(2k+1),-(2k+1)} \hat{H}_1 \ket{(2k+3),-(2k+3)}$
for an odd $N$ and
$\Delta_k=\bra{(2k),-(2k)} \hat{H}_1 \ket{(2k+2),-(2k+2)}$
for an even $N$.

Discussions on connection fields and topological charge densities can 
be carried out, similar to those in the previous section.
After some straight forward calculations, we find:
\beq
{\bf A}=-\frac{\cos\theta}{\sin\theta}\frac{Q_g(|\bf B|)}{|{\bf B}|} 
{\bf e}_\phi.
\eeq
Assuming the field gradient is small, we find that
$Q_g(|{\bf B}|)$ is identical to $Q(|{\bf B}|)$ defined in section III A 
when $B$ is
far away from degeneracy points $B_k$; close to $B_k$ when 
$\gamma |B-B_k|$ are comparable to $\Delta_k$ however,

\beq
Q_g(|{\bf B}|)=S_k + 2|\delta_{k2}|^2
\label{Qg}
\eeq
which varies smoothly from $S_k$ to $S_k+2$.
We want to emphasize that both $\delta_{k1}$ and $\delta_{k2}$ are 
real functions of $|{\bf B}|$ and independent of 
$\theta$ and $\phi$. 

Following discussions in the previous sections,
one obtains the two-form connection fields ${\bf F}_{ab}$ or ${\bf b}$

\beq
{\bf b}=\frac{Q_g(|\bf B|)}{|{\bf B}|^2}
{\bf e}_B+\frac{1}{B}\frac{\partial Q_g(B)}{\partial 
B}\frac{\cos\theta}{\sin\theta} {\bf 
e}_\theta.
\label{bfg}
\eeq

The topological charge density is

\beq
\rho=q_0(N)\delta({\bf B})+\frac{1}{2\pi |{\bf 
B}|^2\sin\theta}
\frac{\partial Q_g(|{\bf B}|)}{\partial B}
[\delta(\theta-\pi)+\delta(\theta)]
\label{seg}
\eeq
Correspondingly,
the surface charge becomes

\beq
\sigma(B)=q_0 (N) \delta(B)+\frac{\partial Q_g(B)}{\partial B}
\label{scg}
\eeq
which is analytical at $B_k$, $k=0,1,...M_N$ because of level 
repulsion. Following Eqs.\ref{co},\ref{Qg}, far away from $B_k$, 
$\sigma(B)$ is vanishingly small. $\sigma(B)$ as a function of $B$ is
numerically plotted in FIG.5.

\begin{figure}
\label{fig:rbph}
\setlength{\unitlength}{1mm}
\begin{picture}(0,0)
\put(40,60){a)}
\put(40,0){b)}
\end{picture}
\epsfig{file=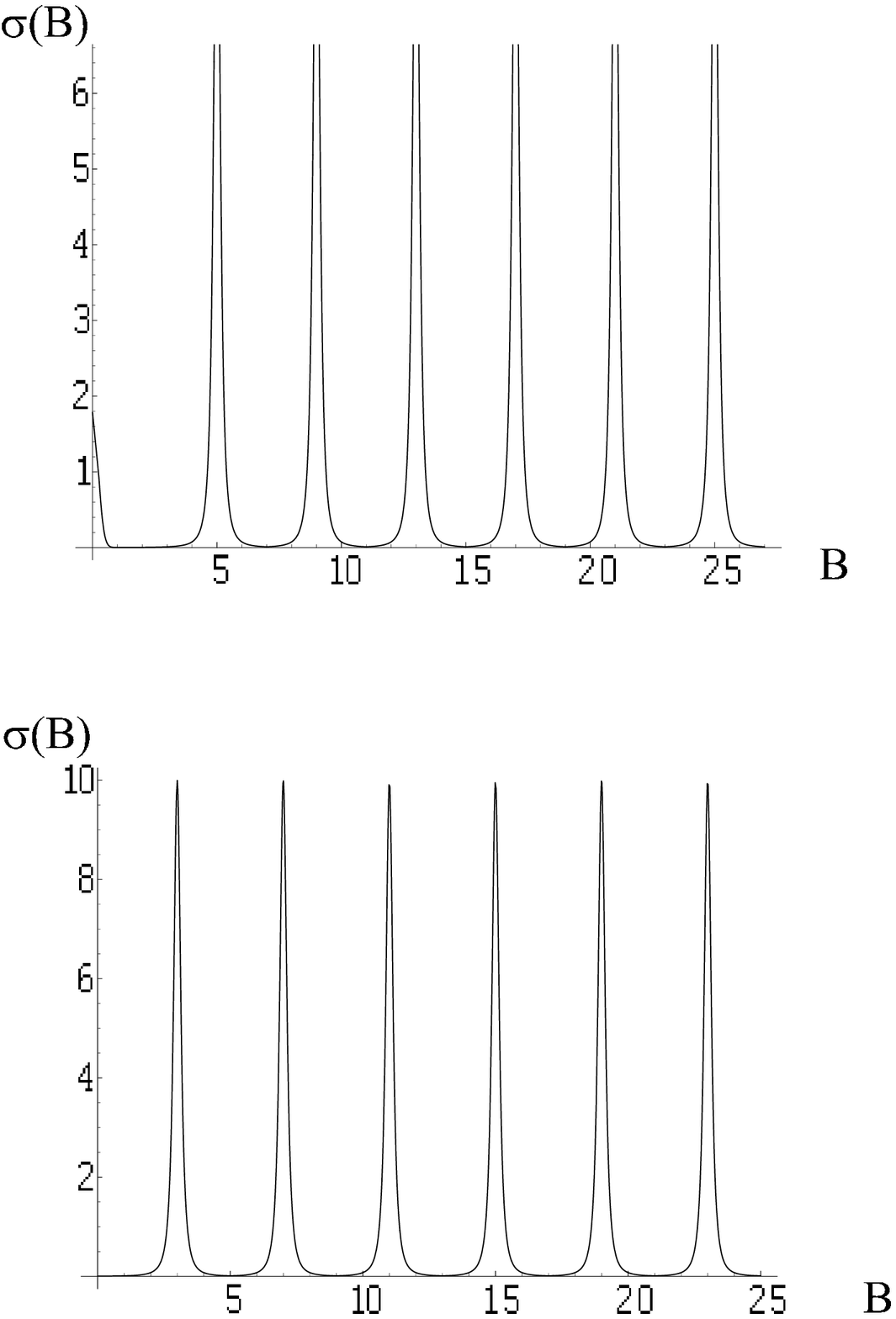,height=120mm,width=80mm}
\caption{The surface charge $\sigma(B)$ as a function of
magnetic field in the presence of
a field gradient $G'=1.32 \mbox{cm}^{-1}$. $B$ is measured in units of
$g_2 \gamma^{-1}$. a) is for an odd number of particles and b)
for an even number of particles.}
\end{figure}

Eqs.\ref{bfg},\ref{scg}
again clearly indicate shell structures.
Indeed, when the width of each shell $2h$ defined before
Eq.\ref{sc} is much 
larger than a characteristic width $W_k$
\beq
2 h \gg  W_k =\frac{\Delta_k}{\gamma},
\eeq
the charge $Q_k$ enclosed in the $k$th shell centered at $B=B_k$ is 

\beq
Q_k=2+O(\frac{W_k^2}{h^2});
\eeq
$Q_k$ approaches two units when the ratio between $2h$ 
and $W_k$ becomes infinity.
And only on these shells, ${\bf b}$ fields have a $\theta$-component.

Therefore,
Eq.\ref{seg} shows that on each shell
the charge distribution is highly anisotropic.
All charges on each shell in this case 
are carried by two monosegments.
Each monosegment carries one unit 
of charge and is centered at ${\bf B}=\pm B_k {\bf e}_z$;
a monosegment is smooth along $z$-axis, the typical width of a
monosegment located at ${\bf B}=\pm B_k{\bf e}_z$ is 
approximately $W_k$.
Topological charges overall distribute
in a chain of monosegments instead of monopoles. 
Finally, in the presence of field gradients,
results in Eq.\ref{conser} are still valid.

Before we leave this subsection, we emphasize that ${\bf b}$-fields 
are gauge invariant under a usual $U(1)$ gauge transformation.
Though this is hardly surprising by virtual of the two-form construction,
in appendix E we nevertheless present an explicit calculation to 
illustrate this 
point in terms of many-body wave functions.

\subsection{Landau-Zener effects}

$B(t)$ varies adiabatically.
When $t \rightarrow -\infty $, 
the ground state is
$\ket{1,-1}$ and when $t \rightarrow +\infty$   
the ground state is $\ket{3,-3}$. Since $B$ changes
adiabatically, for the most of time the system remains in the ground 
state; 
however,
as the change rate is finite,
the system also
makes transitions to an excited state $\ket{1,-1}$ 
at
$t \rightarrow \infty$ with small probability. 

Denoting the excited state 
by $+$ and the
groundstate by $-$, the transition probability can be calculated with
the following formula\cite{QM58}
\beq
W_{+;-} =
 \exp \left[ -2 \; {\cal I}m \int_C^{\tau_0} \! d \tau \,
(E_+(\tau)-E_-(\tau) )
\right]. 
\eeq
Here $\tau_0$ is the complex value at which $E_+(\tau)=E_-(\tau)$ and $C$ 
a
curve from $t \rightarrow - \infty$ to $t \rightarrow + \infty$ passing
above $\tau_0$.
Substituting the results found in Eq.59 into the above expression, we get
\bea
W_{+;-} & = & \exp \left[ -8 \; {\cal I}m \int_0^{i \frac{\Delta_0} 
{v\gamma}} \! \! d \tau \,
\sqrt{\Delta_0^2+ \gamma^2 v^2 \tau^2} \right] \\
&=& \exp \left[-\, 2 \, \frac{\pi}{\gamma v} \Delta_0^2 \right] \nonumber 
\eea
with $v$ as the rate.
\begin{figure}
\label{fig:levcr13}
\epsfig{file=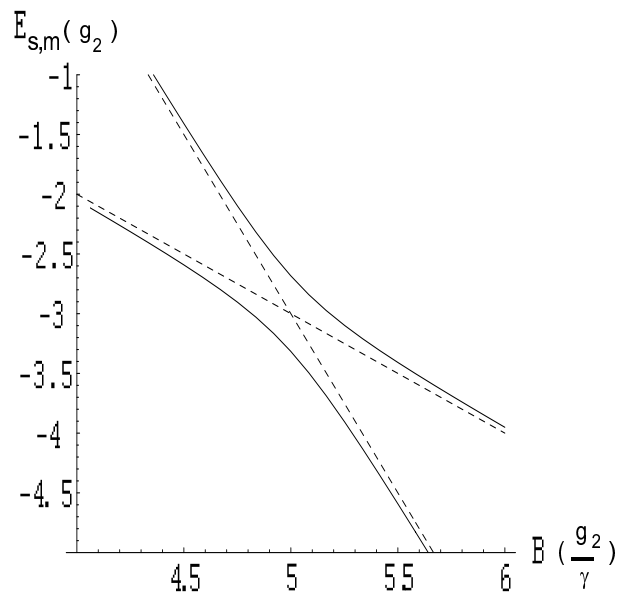,height=80mm,width=80mm}
\caption{Eigenvalues as functions of magnetic fields close
to $B_0$; $G'=1.32 \mbox{ cm}^{-1}$.}
\end{figure}
In FIG.6, we show the energy of these two levels as a
function of magnetic field in the vicinity of
$B_0$.
The steps in the Berry's phases as a function of magnetic field 
become rounded because of level repulsion.

\begin{figure}
\label{fig:rbph}
\setlength{\unitlength}{1mm}
\begin{picture}(0,0)
\put(40,52){a)}
\put(40,-2){b)}
\end{picture}
\epsfig{file=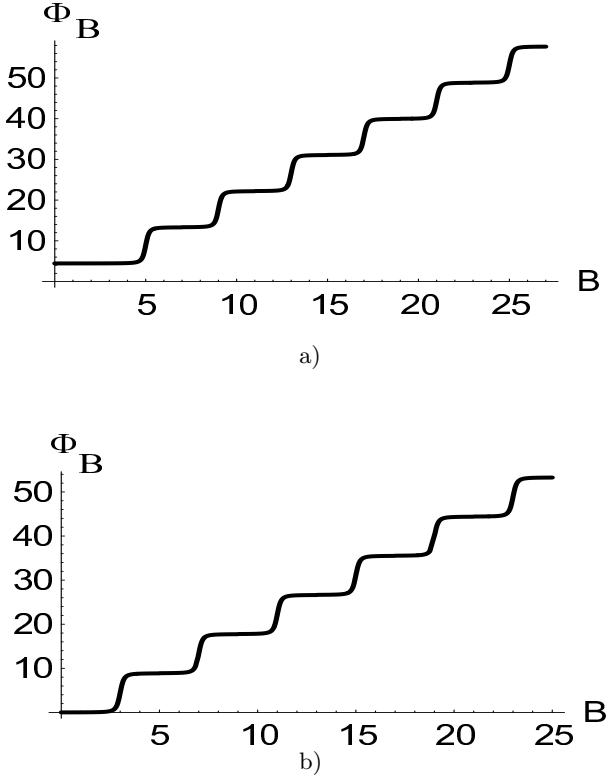,height=100mm,width=80mm}
\caption{Berry's phase in the presence of a field gradient
$G'=1.32 \mbox{cm}^{-1}$. $B$ is measured in units of $g_2 \gamma^{-1}$.
a) is for an odd $N$; b) is for an even $N$.}
\end{figure}

\section{Large $N$ limit}
\subsection{An effective hamiltonian in large $N$ limit}
As proposed in a few previous works, the problem of interacting spin-one 
bosons can be mapped into
a constrained quantum rotor model in the large $N$ limit 
\cite{Zhou01,Zhou01a,Imambekov03,Snoek03,Zhou03}. 
And any microsopic many-body state can be expressed in terms of a wave 
function 
$\psi({\bf n})$ of
a quantum rotor characterized by its direction ${\bf n}$.
In fact, an arbitrary wave function
$\psi({\bf n})$  represents the following microscopic wave function of
a correlated state

\beq
\Psi=
\id {\bf n} \, \psi({\bf n}) \, |{\bf n}>;\, 
|{\bf n}>=\sqrt{\frac{N+1}{2N!}} 
({\bf n}_\alpha \hat{\psi}_\alpha^\dag)^N \ket{0}.
\eeq

Therefore,
a state where ${\bf n}$ is localized on a two sphere corresponds to a polar condensate;
an $S$-wave of ${\bf n}$ represents a rotationally invariant spin singlet
ground state. More discussions about connections between the two
representations can be found in \cite{Zhou03}.

The hamiltonian for spin-one bosons in $|{\bf n}>$- 
representation is given by a quantum rotor model
\beq
H=g_2 \, \hat{{\bf L}}^2 + \gamma \, {\bf B} \cdot \hat{{\bf L}}.
\eeq
The total spin of spin-one particles $\hat{\bf S}_{tot}(=\hat{\bf L})$ is
a differential operator 

\beq
\hat{{\bf L}}= -i \, {\bf n} \times \frac{\partial}{\partial {\bf n}},
\eeq
i.e. the total spin operator is the angular momentum operator defined on the
two-sphere where ${\bf n}$ lives; $\hat{\bf L}$ is also a conjugate 
variable of 
${\bf n}$.
Wave functions further observe the following Ising symmetry

\beq
\psi({\bf 
n})=(-1)^N\psi(-{\bf n})
\label{Ising}
\eeq
for any $N$; this property of 
many-body wave functions 
was identified and 
emphasized in 
previous works on homogeneous gases of spin-one bosons 
\cite{Zhou01,Zhou01a} and on spin-one bosons 
in lattices\cite{Demler02,Imambekov03,Snoek03,Zhou03}.

A direct calculation in Appendix F indicates 
the following relation in the $|{\bf n}>$-representation,

\beq
\hat{\psi}_\alpha^\dag h_{\alpha\beta} \hat{\psi}_\beta
\rightarrow
N \, {\bf n}_\alpha \, h_{\alpha\beta} \, {\bf n}_\beta.
\eeq
Therefore, the field gradient induced quadratic Zeeman term 
discussed in Appendix D
results in 
the following coupling term in the quantum rotor representation (up to a 
shift)
\beq
H_1 \; = N\frac{\hbar^2}{2m} 
h_{\alpha \beta} Q_{\alpha \beta}.
\eeq
Here
\beq
h_{\alpha \beta}({\bf r})=\nabla {\bf \theta}_\eta
\cdot \nabla {\bf \theta}_\xi {\bf S}^{\eta}_{\alpha\gamma} 
{\bf S}^{\xi}_{\gamma\beta};
\eeq
and
$Q_{\alpha \beta}$ is the nematic order parameter
\beq
Q_{\alpha \beta}={\bf n}_{\alpha} \, {\bf n}_{\beta} - \frac{1}{3} 
\delta_{\alpha 
\beta}.
\eeq

For a magnetic field distribution given in this article,
the $h$-matrix is
\beq
h_{\alpha \beta}={G'^2} 
({\bf S}^y {\bf S}^y)_{\alpha \beta}; 
\eeq
therefore, the effective hamiltonian for the gradient term in the large 
$N$ limit is 

\beq
H_1=\frac{\hbar^2 G'^2}{2m} \, N \, \sum_{\alpha\neq y}{\bf 
n}_\alpha^2 
\eeq
which was also proposed in an early paper \cite{Zhou01a}.

\subsection{Effects due to anisotropy}

Taking into account anisotropy, the quantum rotor effective hamiltonian 
for a magnetic field distribution given in Eq.(\ref{fd}) is
\bea
&& H= g_2 \, \hat{{\bf L}}^2 + \gamma \, {\bf B} \cdot \hat{{\bf L}}+
h_{\alpha \, \beta} \, Q_{\alpha \beta} \, N, \nonumber \\
&& h_{\alpha\beta}
=\frac{\hbar^2 G'^2}{2m} \delta_{\alpha\beta}(\delta_{\alpha\, x}
+\delta_{\alpha\, z}).
\eea

The
tunneling term between different total spin states is given by
\bea
&&\bra{s \, m} h_{\alpha \beta} \hat{\psi}^\dag_\alpha \hat{\psi}_\beta 
\ket{s' \, m'}
=  h_{\alpha \beta}
T_{\alpha \beta}(s m,s'm') \, N \nonumber \\
&& T_{\alpha \beta}(sm,s'm')=
\int \! \!
d{\bf n} \; Y^{\ast}_{s m}({\bf n}) {\bf n}_\alpha {\bf n}_\beta \, 
Y_{s'm'}({\bf n}).
\eea

For states $\ket{1,-1}$ and $\ket{3,-3}$,
we have
\beq
T_{\alpha \beta}(3,-3;1,-1) =  \sqrt{\frac{3}{70}}
		\left( \begin{array}{ccc}
                          1 & i & 0\\
			 i &-1 & 0 \\
0& 0 & 0
                  \end{array} \right).
		  \eeq
And the matrix element between $\ket{3,-3}$ and $\ket{1,-1}$ is 

\beq
N \, h_{\alpha \beta} T_{\alpha\beta}(3,-3; {1,-1})
= \frac{\hbar^2 G'^2}{2m} \sqrt{\frac{3}{70}} \, N  
\eeq

Close to $B_0$,
states $\ket{1,-1}$ and $\ket{3,-3}$ are nearly degenerate as 
mentioned before. 
In the two-level subspace, the Hamiltonian is
identical to that in Eq.53; 
$\Delta_0(=h_{\alpha \beta}$
$T_{\alpha \beta})$ is calculated above. 
One can then study the eigen states in this subspace; 
after redefining $\Delta_0$, one obtains results
identical to those in section IV C.

\subsection{Berry's phases}

In the quantum rotor representation, 
the spectra of the Hamiltonian subject to Ising symmetries in 
Eq.\ref{Ising} are
identical to those discussed in section II.
The Berry's phases of many-body ground states in a rotating field
can be evaluated in a straightforward way. All results in section III, IV
can be easily rederived; moreover, this effective description 
gives a simple geometric interpretation of results
obtained in a microscopic calculation. We do not present detailed 
calculations here.


\section{conclusion}

In this article we discuss the Berry's connection fields of many-body 
states of
spin-one bosons with antiferromagnetic interactions. We
show that unlike noninteracting systems, Berry's connection fields
are determined by a linear chain of monopoles; more over in the 
presence of field gradients each monopole becomes a linearly extended 
object
which we call a monosegment. Antiferromagnetic interactions appear to
result in anisotropic expulsion of topological charges from the origin of
parameter space and suppression of Berry's phases in small 
rotating magnetic 
fields.

The peculiar Berry's phases of many-body states might be observed 
by studying resonance transitions between a ground state and a collective 
excitation, similar to NMR experiments carried out earlier\cite{Tycko87}.
In an adiabatically rotating magnetic field, the resonance frequency between
these states should be shifted because of geometric phases. For instance, 
at $B$ less than $B_0$ the shift in the resonance frequency of transitions 
between an excited state
$|2k+1, -2k-1>$ and the ground state $|1,-1>$ for an odd number of 
particles is  

\begin{equation}
\delta \omega =-4k\pi cos\theta_0 \Omega.
\end{equation}
Here $\Omega$ is the rotating frequency and $\theta_0$ is the angle
between the magnetic field and rotation axis $z$ (see section IIIA for 
the geometry). 

{\bf Acknowledgement} One of us (FZ) wants to thank the ASPEN center for 
physics 
for its hospitality
during the year 2003 ASPEN workshop on quantum gases.
He also acknowledges a very pleasant conversation with
S. F. Su, L. T. Wang and Q. Niu 
during which the notion of {\em monoshell} was suggested.
This work was supported by the Foundation FOM, the Netherlands
under contracts 00CCSPP10, 02SIC25 and NWO-MK "projectruimte" 00PR1929.

\appendix

\section{Basic Algebras}

One can verify the following algebra:
\begin{subequations}
\begin{eqnarray}
\lbrack \hat {\bf S}^\alpha, \hpsi_{\beta} \rbrack &=& 
i \epsilon^{\alpha \beta \gamma} \hpsi_{\gamma} \\
\lbrack \hat {\bf S}^\alpha, \hpsi_{\beta}^\dagger \rbrack &=&
i \epsilon^{\alpha \beta \gamma} \hpsi_{\gamma}^\dagger \\
\lbrack \hat {\bf S}^\alpha, \hat {\bf S}^\beta \rbrack &=&
i \epsilon^{\alpha \beta \gamma} \hat {\bf S}^\gamma \\
\nonumber \\
\lbrack \hat \rho, \hpsi_{\alpha} \rbrack &=& - \hpsi_{
\alpha} \\
\lbrack \hat \rho, \hpsi_{\alpha}^\dagger \rbrack
&=&  \hpsi_{\alpha}^\dagger \\
\nonumber \\
\lbrack \hat {\bf S}^\alpha, \hat \rho \rbrack &=& 0
\end{eqnarray}
\end{subequations}
Furthermore, we define the singlet creation operator:
\bea
\hat{A}^\dag &=& \frac{1}{\sqrt{6}} \, \hat{\psi}^\dag_\alpha
\hat{\psi}^\dag_\alpha \mbox{   with $\alpha=x,y,z$} \\
&=& \frac{1}{\sqrt{6}} (\hat{\psi}^\dag_0 \hat{\psi}^\dag_0 
-2 \hat{\psi}^\dag_{-1} \hat{\psi}^\dag_{1}). \nonumber
\eea
This operator satisfies
\beq
\left[ \hat{A}, \hat{A}^\dag \right]=\frac{1}{3}( 3 + 2 \hat{N}).
\eeq

\section{Field gradient and spin conservation}
In this appendix, we prove that a field gradient does not conserve
the total spin of spin-one bosons.
The second quantised form of the total spin operator $\hat{{\bf S}}^2_{tot}$
is given by
\bea
\hat{{\bf S}}^2_{tot} &=& 2 \int \! \! d {\bf x} \, \hat{\psi}^\dag_\alpha({\bf x})
\hat{\psi}_\alpha({\bf x}) \\
 &+& \int \! \! d {\bf x} \int \! \! d {\bf y}
 \hat{\psi}_\alpha^\dag({\bf x}) \hat{\psi}^\dag_\beta ({\bf y}) 
 {\bf S}_{\alpha \alpha'} \cdot
 {\bf S}_{\beta \beta'} \hat{\psi}_{\beta'}({\bf y}) 
\hat{\psi}_{\alpha'}({\bf x}). \nonumber
\eea

We consider the commutator
\beq
\left[ \hat{{\bf S}}^2_{tot} \, , \, \gamma \int \! \! d {\bf x} \,  \, 
{\bf B}({\bf x})
\cdot
\hat{\psi}^\dag_\alpha({\bf x}) {\bf S}_{\alpha \beta} 
\hat{\psi}_\beta({\bf x}) 
\right].
\eeq
Calculations yield
\begin{widetext}
\bea
\lefteqn{\left[ \int \! \! d {\bf x} \, {\bf B}({\bf x}) \cdot 
\hat{\psi}_\alpha^\dag \,
{\bf S}_{\alpha \beta} \hat{\psi}_\beta , \hat{{\bf S}}^2_{tot} \right] =}  \\
&&\left[ \int \! \! d {\bf x} \,\, {\bf B}({\bf x}) \cdot 
\hat{\psi}_\alpha^\dag
{\bf S}_{\alpha \beta} \hat{\psi}_\beta \, , 2 \int \! \! d {\bf x}
\hat{\psi}^\dag_\alpha({\bf x}) \hat{\psi}_\alpha({\bf x}) \right]  \nonumber \\
&+&\left[ \int \! \! d {\bf x} \, {\bf B}({\bf x}) \cdot \hat{\psi}_\alpha^\dag
{\bf S}_{\alpha \beta} \hat{\psi}_\beta \, , \int \! \! d {\bf x} \! 
\int \! \! d {\bf y} \, \,
\hat{\psi}^\dag_\alpha ({\bf x}) \hat{\psi}^\dag_\beta({\bf y}) 
{\bf S}_{\alpha \alpha'} \cdot
{\bf S}_{\beta \beta'} \hat{\psi}_{\beta'}({\bf y}) \hat{\psi}_{\alpha'}({\bf x}) \right]
\nonumber
\eea
\end{widetext}
Using
\beq
\left[ A,BC \right] = \left[ A, B \right]C + B \left[ A,C \right]
\eeq
and
\beq
\left[ \hat{\psi}_\alpha({\bf y}) \, , \hat{\psi}^\dag_\beta({\bf x}) \right] =
\delta_{\alpha \beta} \, \delta({\bf x}-{\bf y})
\eeq
we find that the first term vanishes. The second term can be simplified
using the relations above and in the end we find results in Eq.\ref{ct}.
Now take ${\bf B}({\bf x})$ homogeneous. We find
\bea
\lefteqn{ \left[ \int \! \! d {\bf x} \, {\bf B}({\bf x}) \cdot 
\hat{\psi}_\alpha^\dag
{\bf S}_{\alpha \beta} \hat{\psi}_\beta , \hat{{\bf S}}^2_{tot} \right]=} \\
\lefteqn{ (-2i) {\bf B} \cdot (\int \! \! d {\bf x} \hat{\psi}^\dag_\beta({\bf x}) 
{\bf S}_{\beta \beta'}
\hat{\psi}_{\beta'}  ({\bf x}))} \nonumber \\ 
&\times& (\int \! \! d {\bf y} \hat{\psi}^\dag_\alpha({\bf y})
{\bf S}_{\alpha \alpha'} \hat{\psi}_{\alpha'}({\bf y}) )
- 2 {\bf B} \cdot \int \! \! d {\bf x} \hat{\psi}^\dag_{\alpha}({\bf x})
{\bf S}_{\alpha \beta}
\hat{\psi}_\beta({\bf x}) \nonumber  \\
&=& (-2i) {\bf B} \cdot ( \hat{{\bf S}}_{tot} \times \hat{{\bf S}}_{tot})
-2 {\bf B}
\cdot \hat{{\bf S}}_{tot} \nonumber \\
&=& (-i) B_i \, \epsilon_{ijk} [ \hat{S}^j_{tot} , \hat{S}^k_{tot} ] -2 B^i
\hat{S}^i_{tot} \nonumber \\
&=& 0. \nonumber
\eea
So for a homogeneous magnetic field total spin is conserved and an
inhomogeneous magnetic field breaks conservation of total spin.

\section{Calculations of Berry's connection fields}

\subsection{General discussions: without field gradient}

Define a new coordinate frame $(x',y',z')$ with the $z'$-axis in the
direction of the magnetic field 
${\bf n}=(\sin \theta \cos \phi, \sin \theta \sin \phi,\cos \theta)$, 
and the $x'$-axis in the $(x,y)$-plane.
The creation operator in the local frame is
\beq
\hat{\psi^{''}}_{-1}{}^\dag=-i \, e^{-i \phi} 
\sin^2(\frac{\theta}{2}) 
\hat{\psi}_1^\dag
+i \, \frac{\sin \theta}{\sqrt{2}}  \,\hat{\psi}_0^\dag 
- i \cos^2(\frac{\theta}{2})\,e^{i\phi} \hat{\psi}_{-1}^\dag.
\eeq

Therefore,
$ \bracket{\Psi_{S,-S}({\bf n}) }{d {\Psi_{S,-S}(\bf n)} } $ 
is given by
\begin{widetext}
\bea
\bracket{\Psi_{S,-S}({\bf n})}{d {\Psi_{S,-S}(\bf n)}} 
= C^2 \, S \, \bra{0} \hat{A}^{\frac{N-S}{2}} \hat{\psi^{''}}{}^S_{-1}
(- \, e^{-i \phi} \sin^2(\frac{\theta}{2}) 
\hat{\psi}_1^\dag
+ \,e^{i\phi} \,\cos^2(\frac{\theta}{2}) 
\,\hat{\psi}_{-1}^\dag 
)
(\hat{\psi}_{-1}^{''}{}^\dag){}^{S-1}
\hat{A}^{\dag \frac{N-S}{2}} \ket{0}
\, d \phi \, . \nonumber
\eea
\end{widetext}
The results for connection fields follow this identity.

Now the Berry's phase for $N$ interacting spin-one atoms with total spin 
$S$ 
is given by
\bea
&& \phi_{Berry} =-{\cal I}m \oint_C \bracket{\Psi_{S,-S}({\bf 
n})}{d \Psi_{S,-S}({\bf n})} \nonumber \\
&& = -2 \pi \, S \, \cos(\theta). \nonumber
\eea

\subsection{General discussions: with a field gradient}
Now for $B \sim B_k$
\begin{widetext}
\bea
\bracket{\Psi_{S,-S}({\bf n})}{d \Psi_{S,-S}(\bf n)} &=&
- i \, \cos \theta \left\{ C^2 \, S_k \,  \delta_{k1}^2(B) d \phi \,
\bra{0}
\hat{A}^\frac{N-S_k}{2} \hat{\psi^{''}}_{-1}{}^{S_k}
\hat{\psi^{''}}_{-1}{}^\dag{}^{S_k}
\hat{A}^\dag{}^{\frac{N-1}{2}} \ket{0} \right. \nonumber \\
&+& \left. (S_k+2) \, D^2 \, \delta_{k2}^2(B) \, d \phi \, \bra{0}
\hat{A}^\frac{N-S_k-2}{2} \hat{\psi^{''}}_{-1}{}^{(S_k+2)}
\hat{\psi^{''}}_{-1}{}^\dag{}^{(S_k+2)}
\hat{A}^\dag{}^{\frac{N-S_k-2}{2}} \ket{0} \right. \nonumber \\
&+& \left. (S_k+2) \delta_{k1}(B) \, \delta_{k2}(B) \, C D \, d \phi \,
\bra{0}
\hat{A}^\frac{N-S_k}{2} \hat{\psi^{''}}_{-1}{}^{S_k}
\hat{\psi^{''}}_{-1}{}^\dag{}^{(S_k+2)}
\hat{A}^\dag{}^{\frac{N-S_k-2}{2}} \ket{0} \right. \nonumber \\
&+& \left. S_k \delta_{k1}(B) \, \delta_{k2}(B) \, C D \, d \phi \, \bra{0}
\hat{A}^\frac{N-S_k-2}{2} \hat{\psi^{''}}_{-1}{}^{(S_k+2)}
\hat{\psi^{''}}_{-1}{}^\dag{}^{S_k}
\hat{A}^\dag{}^{\frac{N-S_k}{2}} \ket{0} \right\} \nonumber \\
&=& - (\delta_{k1}^2(B) \, S_k + \delta_{k2}^2(B) \, (S_k+2)) \cos \theta \,
d \phi
\nonumber
\eea
\end{widetext}

\subsection{Berry's local connection fields and topological charge 
densities}

As mentioned in the introduction, Berry's connection fields of the ground 
state are
given 
by
\beq
{\bf A}({\bf B})=-{\cal I}m \, \bra{g}\frac{\partial}{\partial {\bf 
B}}\ket{g}.
\eeq
For a state $|S, -S>$, following discussions in Appendix C1,2, Berry's 
connection potentials in 
spherical coordinates 
$(\theta,\phi,B)$ are
\bea
{\bf A}({\bf B})
= -\frac{Q({\bf B})}{B} \, \cot\theta {\bf e}_{\phi}.
\eea

Now the Berry's connection fields are given by:
\bea
{\bf b} &=& \nabla_{\bf B} \times {\bf A}({\bf B}) \\
&=& \frac{Q(B)}{B^2} {\bf e}_B +
\frac{1}{B} \frac{\partial Q(B)}{\partial B} 
 \cot(\theta)  {\bf e}_\theta.
\eea
For $B\neq 0$, $\theta \neq \pi$ and $\theta \neq 0$, we find
\bea
4 \pi \rho &=& \nabla_{\bf B} \cdot {\bf b} \\
&=& \frac{1}{B^2} \frac{\partial}{\partial B} ( B^2 \frac{Q(B)}{B^2})
+ \frac{1}{B \sin \theta} \frac{\partial}{\partial \theta} 
[ \frac{1}{B} \frac{\partial Q(B)}{\partial B} \cos \theta ] \nonumber \\
&=& \frac{1}{B^2} \frac{\partial Q(B)}{\partial B} -\frac{1}{B^2} 
\frac{\partial Q(B)}{\partial B}=0. \nonumber 
\eea

At $B=0$ or $\theta=\pi$, $\theta=0$ and $B=B_k$ points, the b-fields
are singular. Calculations around those points 
lead to results in 
Eq.\ref{td}.

\section{Quadratic Zeeman effects due to field gradients}

We show explicitly for the invariance of a spin-dependent term in the
interaction. \\ 
Let $U_{\alpha \beta}=\exp [i\mbox{\boldmath $\theta$} \cdot {\bf S}_{\alpha \beta}]$.
\bea
H_{int} &=& \frac{c_2}{2} \int \! \! d {\bf x}  \, \hat{\psi}^\dag_\alpha
\hat{\psi}^\dag_\beta \:
{\bf S}_{\alpha \alpha'} \cdot {\bf S}_{\beta \beta'} \: \hat{\psi}_{\alpha'} 
\hat{\psi}_{\beta'} \\
&=& \frac{c_2}{2} \int \! \! d {\bf x}  \: \hat{\psi}'_\gamma{}^{\dag} \, 
U_{\gamma \alpha}({\bf x}) \,
\hat{\psi}'_\delta{}^{\dag} \, U_{\delta \beta}({\bf x}) \, {\bf S}_{\alpha
\alpha'} \cdot
{\bf S}_{\beta \beta'} \nonumber \\ 
& & U^\dag_{\alpha' \gamma'}({\bf x}) \, \psi'_{\gamma'} \,
U^\dag_{\beta' \delta'}({\bf x}) \, \psi'_{\delta'} \nonumber
\eea
Now using
\bea
\lefteqn{U_{\alpha \alpha'} \, {\bf S}_{\alpha' \beta'} \, 
U^\dag_{\beta' \beta} =} \nonumber \\
&&(U_{\alpha \alpha'} \, {\bf \hat{x}} \cdot
{\bf S}_{\alpha' \beta'} \, U^\dag_{\beta' \beta} , U_{\alpha \alpha'} \, {\bf \hat{y}} \cdot
{\bf S}_{\alpha' \beta'} \, U^\dag_{\beta' \beta} , \nonumber \\
&& U_{\alpha \alpha'} \,
{\bf \hat{z}} \cdot {\bf S}_{\alpha' \beta'} \, U^\dag_{\beta' \beta})
\nonumber \\
&&=( {\bf \hat{x}}' \cdot {\bf S}_{\alpha \beta} \, , {\bf \hat{y}}' \cdot
{\bf S}_{\alpha \beta} \, , {\bf \hat{z}}' \cdot {\bf S}_{\alpha \beta}) 
\eea
where the primes correspond to the rotated coordinate system and the fact
that the inner product is rotationally invariant we find that the
spin-dependent interaction term is invariant under the local spin rotation.

Now the kinetic term
\beq
H=\int \! \! d {\bf x} \, -\frac{\hbar^2}{2 m} \hat{\psi}^\dag_\alpha({\bf x})
\mbox{\boldmath $\nabla$}^2 \hat{\psi}_\alpha({\bf x})
\eeq
becomes in terms of $\hat{\psi}'{}^{\dag}_\alpha$ and $\hat{\psi}'_\beta$
\bea
\lefteqn{H = \int \! \! d {\bf x} \, \hat{\psi}'{}^{\dag}_\delta \,
U_{\delta \epsilon}({\bf x}) \,
\mbox{\boldmath $\nabla$}^2 \,
U^\dag_{\epsilon \alpha}({\bf x}) \,
\hat{\psi}'_\alpha({\bf x})} \\
&=& \int \! \! d {\bf x} \, -\frac{\hbar^2}{2m} \, \hat{\psi}'{}^\dag_\alpha({\bf x})
\, \mbox{\boldmath $\nabla$}^2 \, \hat{\psi}'{}^\dag_\alpha({\bf x}) \nonumber \\
&+& \int \! \! d {\bf x} \, -\frac{\hbar^2}{2m} \,
\hat{\psi}'{}^\dag_\alpha({\bf x}) \, U_{\alpha \beta} \, (\mbox{\boldmath $\nabla$} \,
U^\dag_{\beta \gamma}) \cdot
\mbox{\boldmath $\nabla$} \hat{\psi}'_\gamma({\bf x}) \nonumber \\
&+& \int \! \! d {\bf x} \, -\frac{\hbar^2}{2m}  \hat{\psi}'{}^\dag_\alpha({\bf x}) (
U_{\alpha \beta}({\bf x}) \,
\mbox{\boldmath $\nabla$}^2 U^\dag_{\beta \gamma}({\bf x})) \, 
\hat{\psi}'_\gamma ({\bf x}) \nonumber
\eea

Using the relation
\beq
\frac{\partial}{\partial x_i} \exp \left[ i \mbox{\boldmath $\theta$} ({\bf x}) \cdot
\hat{{\bf S}} \right] = i \, ( \frac{\partial \mbox{\boldmath $\theta$}({\bf x})}{\partial
x_i} \cdot \hat{{\bf S}}) \,
\exp
\left[ i \mbox{\boldmath $\theta$} \cdot \hat{{\bf S}} \right]
\eeq

and 
\beq
\frac{\partial \mbox{\boldmath $\theta$}}{\partial x_i}=\delta_{i,x} 
G' \, {\bf \hat{y}}
\eeq
for the ansatz in Eq.\ref{fd},
we have $H=H_0+H_1$, and
\bea
H_1 &=& \frac{\hbar^2}{2m}  \int \! \! d {\bf x} \,
\hat{\psi}'{}^{\dag}_\alpha U_{\alpha \beta}( i G'  
S^y)_{\beta\gamma} 
U^\dag_{\gamma \eta}
\frac{\partial}{\partial x} \hat{\psi}'_\eta  \\
&-& \frac{\hbar^2}{2m} \int \! \! d {\bf x} \, \hat{\psi}'{}^{\dag}_\alpha 
U_{\alpha \beta} [-G'^2
(S^y)^2]_{\beta \gamma} U^\dag_{\gamma \eta} \hat{\psi}'_\eta. 
\nonumber
\eea

By keeping only the zero mode contributions, we
find

\beq
H_1= \frac{\hbar^2 G'^2 }{2m} \, \hat{\psi}^\dag_\alpha \,
(S^y)^2_{\alpha \beta}
\, \hat{\psi}_\beta.
\eeq
This is the effective quadratic coupling in the presence of a field 
gradient which was previously obtained in Ref.\cite{Ho00}.

\section{Gauge invariance of topological fields}

\subsection{General}

We define the following basis
\begin{widetext}
\bea
\ket{{S,-S};{\bf n}}= C(S) \, e^{i \, P(B,\theta,\phi)} \,
\left[ -i \sin^2(\frac{\theta}{2}) \, e^{-i \phi} \, \hat{\psi}_{+1}^\dag
+ \frac{i}{\sqrt{2}} \sin \theta \,
\,  \hat{\psi}_0^\dag - i \cos^2(\frac{\theta}{2}) \, e^{i \phi} \,
\hat{\psi}_{-1}^{\dag}
\right]^S
\hat{A}^\dag{}^\frac{N-S}{2} \ket{0}.
\eea
\end{widetext}

Close to $B_k$ points,
ground states are in general superpositions of these states;
using the ansatz introduced in section IV D, we find

\begin{widetext}
\begin{equation}
\Psi(B,{\bf n})= \delta_{k1}(B) 
\ket{S_k,-S_k;{\bf n}} + \delta_{k2}(B) \, e^{i d_{k}(\theta,\phi,B)} 
\ket{S_k+2,-S_k-2; {\bf n}}
\end{equation}
\end{widetext}
with $\delta_{k1,k2}$, $d_{k}$ functions of $(B,\phi,\theta)$.
Calculations yield $\delta_{k1,k2}$ as given in section IV D; 
$d_k=0$.

Therefore we find

\bea
&& {\bf A}({\bf B})=-\frac{\partial P}{\partial B} {\bf e}_B  
-\frac{1}{B} \frac{\partial P}{\partial \theta} {\bf e}_\theta 
\nonumber \\
&& -\frac{1}{B \sin \theta} 
(\frac{\partial P}{\partial \phi}+Q_g(B) \cos \theta) {\bf e}_\phi
\eea
Note that Berry's connection potentials ${\bf A}({\bf B})$
transform in an expected form
\beq
{\bf A}({\bf B})\rightarrow {\bf A}({\bf B})+ {\bf \nabla} \, 
\delta P(B,\theta,\phi),
\eeq
under a gauge transformtion $P\rightarrow P +\delta P$.

\bea
&& {\bf b}({\bf B}) = \frac{Q_g(B)}{B^2} {\bf e}_B + \frac{1}{B} 
\frac{\partial Q_g(B)}{\partial B} \frac{\cos \theta}{\sin \theta} 
{\bf e}_\theta
\eea
We find on the k-th shell two monopoles, one at the southern and one at
the northern pole.
The connection field ${\bf b}({\bf B})$ does not depend on the choice of
$P(B,\theta,\phi)$ i.e. the gauge.

\subsection{Case $P(B,\theta,\phi)=0$}

We find in this limit,
\beq
{\bf A}({\bf B})=
 -\frac{\cos\theta}{B \sin \theta}
Q_g(B) {\bf e}_\phi
\eeq

\bea
&& {\bf b}({\bf B})= \frac{Q_g(B)}{B^2} {\bf e}_B + \frac{1}{B}
\frac{\partial Q_g(B)}{\partial B} \frac{\cos \theta}{\sin \theta}
{\bf e}_\theta
\eea

\section{Estimate for cold sodium atoms}
Consider a trap with $N=10^6$ atoms,
a density $n=10^{14} cm^{-3}$ and
$a_2-a_0=0.32 \, \mbox{nm}$. We have $g_F=-\frac{1}{2}$ for $^{23}$Na.
The Zeeman effect is characterized by $\gamma$ 
\bea
\gamma=g_F \, \mu_B &=& -3.3585 \, \cdot \, 10^{-5} \, \frac{K}{G}. 
\eea
Note that $\gamma$ is negative. Previous results can be applied provided we
take for the magnetic field ${\bf B}= - B {\bf n}(\theta,\phi)$. 
One also finds
\beq
\Delta_0 = 2.18 \, \cdot \,  10^{-15} \, K \, m^2 \, {G'}^2 \, , 
\eeq
where we have the following units: $G'$ in $m^{-1}$ and $\Delta$ in $K$.
Furthermore
\beq
g_2=10^{-15} \, K.
\eeq
The eigenvalues of $H$ in the vicinity of $B_0$ are
\bea
\lefteqn{E^{\pm}(\delta B)=-3 \, g_2 -2 \gamma \delta B}  \\
&& \pm \gamma \sqrt{4.23 \,
\cdot \, 10^{-21}
G^2 \, m^4 \, G'^4 \,+ (\delta B)^2}. \nonumber
\eea
The relative level repulsion $\frac{\Delta E}{g_2}$ is
$\sim 4 \, m^2 \, G'^2$ with $G'$ in $m^{-1}$.

\section{Calculation of matrix elements in a large-N limit}

Now the general operator 
$\hat{\psi}_\alpha^\dag h_{\alpha \beta} \hat{\psi}_\beta$ with 
$h_{\alpha \beta}$ some matrix 
acts on the wave function as follows:
\begin{widetext}
\bea
\langle \psi_1({\bf n})|
\hat{\psi}^\dag_\alpha 
h_{\alpha\beta} \hat{\psi}_\beta
| \psi_2({\bf n}) \rangle &=& \id {\bf n}_1 \id
{\bf n}_2  \, \frac{N+1}{2N!} \, \bra{0}(n^1_\gamma \hat{\psi}_\gamma)^N
\hat{\psi}_\alpha^\dag h_{\alpha \beta}
\hat{\psi}_\beta (n^2_\delta
\hat{\psi}_\delta^\dag)^N \ket{0} \, \psi^\ast_1({\bf n}_1) \psi_2({\bf n}_2) \\
&=& \id {\bf n}_1 \id {\bf n}_2\,\frac{N+1}{2 N!} \, N^2
n^1_\alpha h_{\alpha \beta} n^2_\beta \, \psi_1^\ast({\bf n}_1) 
\psi_2({\bf n}_2) \,
\bra{0}(n^1_\gamma \hat{\psi}_\gamma)^{N-1} 
(n^2_\delta \hat{\psi}_\delta^\dag)^{N-1} \ket{0} \nonumber \\
&=& \id {\bf n} \, N \, n_\alpha h_{\alpha \beta} n_\beta \,
\psi_1^\ast({\bf n}) \psi_2({\bf n}). \nonumber
\eea
\end{widetext}

\end{document}